\documentclass[aps,pra,showpacs,twocolumn,superscriptaddress,longbibliography]{revtex4-1}
\usepackage{graphicx}
\usepackage[usenames]{color}
\usepackage{amssymb,amsmath}
\usepackage{siunitx}
\usepackage{xcolor}
\usepackage{setspace} 
\usepackage{float} 
\usepackage{tabularx}
\usepackage{booktabs}

\newcommand{\be}[1]{\begin{equation}\label{#1}}
\newcommand{\ee}{\end{equation}}

\usepackage[LGRgreek]{mathastext}
\usepackage{lipsum}

\newcommand{\gj}[6]{ \begin{pmatrix}
 #1 & #2 & #3 \\
 #4 & #5 & #6 
 \end{pmatrix}}

\usepackage{graphicx}
\usepackage{dcolumn}
\usepackage{bm}
\usepackage{hyperref}
\usepackage{titlesec}
\usepackage{lipsum}
\usepackage{pbox}

\begin{document}

\title{Sequential single-photon and direct two-photon absorption processes for Xe interacting with attosecond XUV pulses}

\author{A. Hadjipittas}
\affiliation{Department of Physics and Astronomy, University College London, Gower Street, London WC1E 6BT, United Kingdom}


\author{H. I. B. Banks}
\affiliation{Department of Physics and Astronomy, University College London, Gower Street, London WC1E 6BT, United Kingdom}

\author{B. Bergues}
\affiliation{Department of Physics, Ludwig-Maximilians-Universit\"{a}t Munich, Am Coulombwall 1, D-85748 Garching, Germany}
\affiliation{Max Planck Institute of Quantum Optics, Hans-Kopfermann-Str. 1, D-85748 Garching, Germany}

\author{A. Emmanouilidou}
\affiliation{Department of Physics and Astronomy, University College London, Gower Street, London WC1E 6BT, United Kingdom}



\begin{abstract}

We investigate the interaction of Xe with isolated attosecond XUV pulses. Specifically, we calculate the ion yields and determine the pathways leading to the formation of ionic charged states up to Xe$^{5+}$. To do so, in our formulation we account for single-photon absorption, sequential multi-photon absorption, direct two-photon absorption, single and double Auger decays, and shake-off. We compare our results for the ion yields and for ion yield ratios with recent experimental results obtained for 93 eV and 115 eV attosecond XUV pulses. In particular, we investigate the role that a sequence of two single-photon ionization processes plays in the formation of Xe$^{4+}$. We find that each one of these two processes ionizes a core electron and thus leads to the formation of a double core-hole state. Remarkably, we find that  the formation of   Xe$^{5+}$ involves a  direct two-photon absorption process and the absorption of a total of three photons.
  
 \end{abstract}
 


\date{\today}

\maketitle
\section{Introduction}

The advent of free electron lasers (FEL) \cite{Pellegrini2012}, has allowed for the production of ultra-short and high-energy laser pulses. These XUV pulses allow the ionization of inner-bound electrons that trigger a plethora of processes in atoms and molecules \cite{Marangos_2011, Ullrich_2012, Wallis2015}. Xenon, with 54 electrons, is an ideal atom to investigate the effect that different ionization processes have on the formation of highly charged  ionic states \cite{Pi_XeStudy, Costello_XeStudy, Amusia_XeStudy,  Son_2012, Toyota_2017, Rudek_2012, Santra_Xe}. Previous studies have investigated the formation of Xe ion states up to $Xe^{21+}$ \cite{Makris_21, Lambropoulos_21, Sorokin_21} when a pulse of femtosecond duration at 93 eV interacts with Xe. While FEL sources deliver high XUV pulse energies, the pulse duration is typically limited to the femtosecond range. In contrast, high-harmonic generation (HHG) based XUV sources can deliver isolated  attosecond XUV pulses but the output pulse energy is limited by the low infrared to XUV conversion efficiency of the HHG process. This has prevented the observation of attosecond multi-photon interactions with inner-shell electrons for a long time. Such attosecond interactions were observed experimentally in Xe only recently \cite{Bergues}. The results of that study exhibited strong deviations with respect to sequential ionization via ionic ground states \cite{Bergues}, which dominates the formation of lower-charged ionic states for femtosecond  pulses \cite{Makris_21, Lambropoulos_21}. Hence, the prevalent pathways for the formation of Xe ion charged states  in the attosecond regime is still an open question.

Here, we address this question  and model the interaction of Xe with an attosecond XUV pulse of energy 93 eV and 115 eV. The pulse parameters that we consider are chosen so that we can directly compare our results for ion yields up to $Xe^{5+}$ and our results for ratios of the ion yields with the experimental ones obtained in Ref. \cite{Bergues}. Specifically, the pulses considered in Ref. \cite{Bergues} have photon energies of 93 eV and 115 eV and a duration of about 340 attoseconds (as). 
Unlike previous studies \cite{Bergues}, we account for sequential single-photon absorption processes via the creation of multiple core hole states \cite{MultipleCoreHoles1, MultipleCoreHoles2}. Moreover, we account for single-electron ionization by a two-photon absorption process, referred to as direct two-photon process \cite{Lambropoulos_21, Nikolopoulos_2011}. This latter process has been found to affect the formation of ion charged states above Xe$^{7+}$ in Ref. \cite{Lambropoulos_21}, where an XUV pulse of femtosecond duration is considered. 

Pulses with photon energy of 93 eV and 115 eV can access and ionize electrons from the 4d sub-shell. The processes considered in  our model include   a single electron ionization by single-photon absorption or by a direct two-photon absorption. In addition, we account for Auger decays \cite{Auger}. In an Auger process, an electron falls from a higher-energy shell filling in an inner-shell hole. The energy released leads to the ionization of one or two bound electrons. We refer to the Auger decay as single or double depending on whether it leads to the ionization of one or two bound electrons, respectively. We also account for shake-off processes \cite{shake-off}, resulting in the escape of a second electron following an ionization by a  single-photon process.

In section $II$, we describe the method that we use  to investigate the interaction of Xe with an attosecond XUV pulse. In particular, we describe how to obtain the single-photon ionization cross sections and Auger decay rates that are involved in the rate equations \cite{Makris_21, Santra_RE_2007} that we employ. In section $III$, we compute ion yields and  yield ratios and compare them with experimental results \cite{Bergues}. In particular, we identify the main pathways leading to the formation of  charged states up to Xe$^{5+}$.

 \section{Method}
 
 We employ rate equations, as in Ref.\cite{Wallis2014} but with additional processes, in order to obtain the yields and pathways of the final ion states. In the rate equations we consider terms involving single-photon and two-photon ionization transitions, the Auger and double Auger decays as well as shake-off processes. The electronic configuration of Xe is $1s^22s^22p^63s^23p^63d^{10}4s^24p^64d^{10}5s^25p^6$. A shell is distinguished by the $n$ quantum number and a sub-shell by the $n,l$ quantum numbers. A sub-shell is made up of $2l+1$ orbitals, where each orbital has an occupancy of 0, 1 or 2 electrons. In each $np$ sub-shell we consider the orbitals $np_x$, $np_y$ and $np_z$, and in each $nd$ sub-shell weions consider the orbitals $nd_{xy}$, $nd_{yz}$, $nd_{xz}$, $nd_{x^2-y^2}$ and $nd_{z^2}$.

\subsection{Bound and continuum orbitals}
We denote the bound orbital wavefunction as $\phi_{i}$ and the continuum orbital wavefunction as $\phi_{\epsilon,l',m'}$. To calculate the bound orbital wavefunctions, we use the molecular computing package Molpro \cite{molpro} with the augmented quadruple-zeta plus polarization (AQZP) basis set \cite{AQZP}. This basis set expresses the orbitals as a combination of $l,m$ quantum numbers, whereas in our previous studies of Ar \cite{Wallis2014, Wallis2015}, each orbital was expressed by well-defined $l,m$ numbers and the 6-311G basis set was employed. Specifically, we express the bound orbital wavefunction as a product of a radial component and a spherical harmonic $Y_{l,m}(\theta,\phi)$ as follows

\begin{equation}
\label{eq:SCE}
\phi_{i}(r) = \sum_{l,m} P_{i,l,m}(r)Y_{l,m}(\theta,\phi)/r.
\end{equation}

\noindent To calculate the continuum wavefunction, we use the Herman-Skillman code \cite{HermanSkillman1, HermanSkillman2} to obtain the Hartree-Fock-Slater potential and the Numerov method \cite{Numerov} to obtain the radial part of the wavefunction, as was done in our previous works \cite{Wallis2014, Hadjipittas_2019}. By multiplying the radial part with a spherical harmonic, the continuum wavefunction is given as

\begin{equation}
\label{eq:SCE_continuum}
\phi_{\epsilon,l',m'}(r) = P_{\epsilon,l'}(r)Y_{l',m'}(\theta,\phi)/r.
\end{equation}

\noindent By expressing the bound and continuum orbitals as a product of a radial and an angular component, see Eq. (\ref{eq:SCE}) and Eq. (\ref{eq:SCE_continuum}), we significantly simplify the evaluation of the single-photon ionization cross sections and the Auger rates, see sections \ref{sec:Photo} and \ref{sec:Auger}.

\subsection{Single-photon ionization cross sections}
\label{sec:Photo}

In order to calculate the photo-ionization cross section for an electron to transition from the bound orbital $\phi_{i}$ to the continuum orbital $\phi_{\epsilon,l',m'}$, we use the equation below \cite{Sakurai}

\begin{equation}
\label{photoionisation}
\sigma_{i \rightarrow \epsilon,l',m'} = \frac{4}{3} \alpha \pi^2 \omega N_i \sum_{M=-1,0,1} | D^M_{i \rightarrow \epsilon,l',m'} |^2,
\end{equation}

\noindent where $\alpha$ is the fine structure constant, $N_i$ is the number of electrons in the initial orbital $i$, $\omega$ is the photon energy and $M$ is the polarization of the photon. The matrix element $D^M_{i \rightarrow \epsilon,l',m'}$ is given by

\begin{equation}
\label{MatrixElement}
D^M_{i \rightarrow \epsilon,l',m'} = \int \phi_{i} (r) \phi^*_{\epsilon,l',m'}(r)\sqrt{\frac{4 \pi}{3}}r Y_{1M}(\theta,\phi)dr.
\end{equation}

\noindent Subtituting in Eq. (\ref{MatrixElement}), the expansion for the bound and continuum orbitals from Eq. (\ref{eq:SCE}), we obtain the following: 

\begin{equation}
\begin{split}
D^M_{i \rightarrow \epsilon,l',m'} = \sqrt{\dfrac{4\pi}{3}} \sum_{lm} \int_{0}^{ \infty }dr P_{i,l,m}(r) r P_{\epsilon,l'}(r) \\ \times \int d\Omega Y_{l,m}(\theta,\phi) Y^*_{l',m'}(\theta, \phi) Y_{1M}(\theta, \phi).
\end{split}
\label{ProFinalMatrixEquation}
\end{equation}

\noindent Next, we calculate the angular integrals in terms of the Weigner-3j symbols \cite{Weigner_3j} and obtain

\begin{equation}
\begin{split}
D^M_{i \rightarrow \epsilon,l',m'} = \sum_{lm}(-1)^{m^{'}}\sqrt{(2l+1)(2l'+1)} \\ \times\begin{pmatrix}
l' & l & 1\\
0 & 0 & 0
\end{pmatrix}
\begin{pmatrix}
l' & l & 1\\
-m' & m & M
\end{pmatrix} \\
\times \int^\infty_0 dr P_{i,l,m}(r)rP_{\epsilon,l'}(r).
\end{split}
\label{FinalMatrixElement}
\end{equation}

\noindent Since only the energy of the final continuum orbital is of relevance, we sum in Eq. (\ref{photoionisation}) over all $l'$ and $m'$ numbers to obtain

\begin{equation}
\label{photoionisation_SumOverContinuum}
\sigma_{i \rightarrow \epsilon} = \frac{4}{3} \alpha \pi^2 \omega N_i  \sum_{l',m'}  \sum_{M=-1,0,1} | D^M_{i \rightarrow \epsilon,l',m'} |^2.
\end{equation}

\noindent In our calculations, the electronic configurations of Xe involved in the rate equations are expressed in terms of sub-shells. Hence, to find the single-photon ionization cross section from a certain sub-shell, we have to sum over all the cross sections involving the orbitals in this sub-shell. For instance $\sigma_{2_p} = \sigma_{2_{p_{x}}}  + \sigma_{2_{p_{y}}}  + \sigma_{2_{p_{z}}} $, where each of the $\sigma_{2_{p_{x}}}$ , $\sigma_{2_{p_{y}}}$, $\sigma_{2_{p_{z}}}$ are computed using Eq. (\ref{photoionisation_SumOverContinuum}).

 \begin{table}
\caption{\label{tab:XsectionComparisonTable}Single-photon ionization cross sections of neutral Xe interacting with a pulse of 93 eV photon energy. The units of the cross sections are $cm^2$.} \begin{ruledtabular}
\begin{tabular}{lllll}
\centering   & Ref.\cite{Becker}  & Ref. \cite{Holland_1979} & Ref.\cite{Yeh} & This work\tabularnewline
\hline
$\sigma_{n=5}$ & 1.64$\times10^{-18}$ & 1.47$\times10^{-18}$ & 3.61$\times10^{-19}$ & 3.30$\times10^{-19}$ \\
$ \sigma_{4d^{10}}$ & - & - & 1.71$\times10^{-17}$ & 2.24$\times10^{-17}$ \\
\end{tabular}
\end{ruledtabular}
\end{table}

 In Table \ref{tab:XsectionComparisonTable}, we compare our results with previous theoretical \cite{Yeh} and experimental \cite{Becker, Holland_1979}   single-photon ionization cross sections.

We find that our computed cross sections for single-photon ionization from the $4d$ sub-shell,  $\sigma_{4d^{10}}$, and from the $n=5$ shell, $\sigma_{n=5}$,  are in very good agreement with the theoretical results in Ref. \cite{Yeh}. The difference between our work and Ref. \cite{Yeh} is that the latter employs the Hartree-Fock-Slater method to obtain both the bound and continuum orbitals, while we compute more accurately the bound orbitals using Molpro. 

Moreover, we find that our cross section for ionization from the valence orbitals, $\sigma_{n=5}$, is roughly four times smaller than the one obtained experimentally \cite{Becker, Holland_1979}. This is an accord with Ref. \cite{Amusia_1973} where it is explained that single-particle approximations lead to smaller computed valence cross sections compared to experimental ones. Since our valence cross sections differ from the experimental ones, we obtain results using our computed valence cross sections as well as using the experimental valence cross sections. We find that both sets of cross section provide very similar results for the ion yields and the prevalent pathways. Therefore, in what follows, we present the results obtained using our computed valence ionization cross sections.

\subsection{Two-photon ionization cross sections}
Two-photon ionization involves a single electron ionization following the simultaneous absorption of two photons. The two-photon ionization cross sections are computed via a method of scaling \cite{Lambropoulos_Scaling} and are the ones considered in Ref. \cite{Lambropoulos_21}. For the long pulse considered in Ref. \cite{Lambropoulos_21}, two-photon ionization processes are included for transitions starting from Xe ion states with charge 5 and higher. For the short pulse employed in our work, we consider all two-photon ionization processes that are energetically allowed. However, if for a certain transition, both a single-photon and a two-photon ionization process are energetically allowed, we only account for the single-photon one. The reason is that the single-photon ionization cross sections is roughly thirty orders of magnitude larger than the two-photon ionization cross section. Given the values for the two-photon ionization cross sections obtained in Ref. \cite{Lambropoulos_21}, we estimate that the two-photon ionization cross sections considered in our work vary between $10^{-48}$ $cm^4s$ and $10^{-47}$ $cm^4s$. We obtain two different sets of results, one set using $10^{-47}$ $cm^4s$ for all two-photon ionization cross sections and one using $10^{-48}$ $cm^4s$. We find that both cross section values result in similar pathways. However, the value of $10^{-47}$ $cm^4s$ for the two-photon ionization cross sections leads to a better agreement with the experimental results for Xe$^{5+}$. Thus, the results presented in section $III$, are for a two-photon ionization cross section of $10^{-47}$ $cm^4s$.

\subsection{Auger Decay}
\label{sec:Auger}
\noindent The Auger rate is defined as follows \cite{Fermi}

\begin{equation}
\Gamma=\overline{\sum}2\pi |\mathcal{M}|^2\equiv\overline{\sum}2\pi |\langle\Psi_{fi}|H_I|\Psi_{in}\rangle|^2,
\label{eq:Auger_General}
\end{equation}

\noindent where $\overline{\sum}$ means a summation over final states and an average over the initials states. The operator $H_I$ describes the Coulomb repulsion between the two electrons involved in the Auger transition. The derivation of the Auger decay rate for molecules in our previous work \cite{Emmanoulidou_Banks_Tennyson_Auger} involves bound molecular orbitals which are expressed as a sum of $l,m$ quantum numbers. In contrast, our previous work regarding the interaction of free-electron laser pulses with Ar \cite{Wallis2014, Wallis2015} involves bound orbitals, where only one $l$ quantum number is associated with each orbital. Since, for Xe we consider bound orbitals which are expressed as a sum of $l,m$ quantum numbers, we adapt our formulation of the Auger process for molecules to atoms. As a result, we find that the matrix element for the Auger rate involving two valence orbitals $a$ and $b$, an inner-shell orbital $c$ and a continuum orbital $\epsilon$ with quantum numbers $l',m'$ to be

\begin{equation}
\label{Auger_MatrixElement}
\begin{aligned}
&\mathcal{M}=\delta_{S',S}\delta_{M',M}\sum_{\substack{ l_c,m_c,k\\l_a,m_a,l_b,m_b}}\sum_{q=-k}^{k}\int{dr_1}\int{dr_2}\\
&(-1)^{m+m_c+q}\sqrt{(2l'+1)(2l_c+1)(2l_b+1)(2l_a+1)}\\
&\left[ P_{\epsilon,l'}(r_1) P_{c,l_c,m_c}(r_2)\frac{r^k_<}{r^{k+1}_>}P_{b,l_b,m_b}(r_1)P_{a,l_a,m_a}(r_2) 
\vphantom{\gj{l_c}{k}{l_a}{0}{0}{0}\gj{l_c}{k}{l_a}{-m_c}{q}{m_a}}
\right.\\
&\left.
\gj{l'}{k}{l_b}{0}{0}{0}\gj{l'}{k}{l_b}{-m}{-q}{m_b}
\gj{l_c}{k}{l_a}{0}{0}{0}\gj{l_c}{k}{l_a}{-m_c}{q}{m_a}\right.\\
&+(-1)^S\left. P_{\epsilon,l'}(r_1) P_{c,l_c,m_c}(r_2)\frac{r^k_<}{r^{k+1}_>}P_{a,l_a,m_a}(r_1)P_{b,l_b,m_b}(r_2)
\vphantom{\gj{l_c}{k}{l_a}{0}{0}{0}\gj{l_c}{k}{l_a}{-m_c}{q}{m_a}}
\right.\\
&\left.
\gj{l'}{k}{l_a}{0}{0}{0}\gj{l'}{k}{l_a}{-m}{-q}{m_a}
\gj{l_c}{k}{l_b}{0}{0}{0}\gj{l_c}{k}{l_b}{-m_c}{q}{m_b}\right],
\end{aligned}
\end{equation}

\noindent where $r_<=\min(r_1,r_2)$ and $r_>=\max(r_1,r_2)$. The values $k$ and $q$ are the angular and magnetic quantum numbers of the spherical harmonics involved in the multipole expansion of the Coulomb interaction term $1/r_{12}$. $S$, $S'$, $M_S$ and $M'_S$ are the initial and final total spins and the projection of these spins. The equation for the total Auger rate is given by Eq. (\ref{Auger}). 

\begin{equation}
\Gamma_{b,a \rightarrow c} = \sum_{S,M_S,S',M'_S} \pi N_{ab} N_h \times \sum_{\l',m'} \vert M\vert^2 ,
\label{Auger}
\end{equation}

\noindent where $N_h$ is the number of core holes in orbital $c$ and $N_{ab}$ is a normalisation factor given by 

\begin{align*}
N_{ab} &= \dfrac{N_aN_b}{2 \times 2} \;\;\;\;\;\;\;\text{valence electrons in different orbitals}, \\ & =\dfrac{N_a(N_a-1)}{2 \times 2 \times 1} \;\;\;\;\;\text{valence electrons in the same orbital},
\end{align*}

\noindent where $N_a$ and $N_b$ denote the occupation numbers of orbitals $a$ and $b$. In order to obtain the Auger rate $\Gamma_{s,t \rightarrow u,\epsilon}$ between sub-shells $s$, $t$ and $u$, we add the Auger rates $\Gamma_{b,a \rightarrow c,\epsilon}$ over the $a$ and $b$ orbitals in the $s$, $t$ sub-shells. However, we do not sum over the $c$ orbitals in sub-shell $u$, since we average over the initial states. 

\subsection{Double Auger Decay}
The only energetically allowed double Auger decay process involves $Xe^+$ with a $4d$ hole. In the double Auger process a $5p$ electron drops in to fill the $4d$ hole, while two more $5p$ electrons escape to the continuum. According to  Ref. \cite{Becker} the double Auger decay rate is equal to 21\% of the single Auger decay rate that involves the same initial state as the double Auger decay. The single Auger processes involve either a $5p$ electron filling in the $4d$ hole and the ionization of a $5p$ electron or a $5s$ electron filling in the $4d$ hole while  a  $5p$ or a $5s$ electron escapes. We find that the value of the double Auger decay rate is $6.14 \times 10^{-4}$ a.u.

\subsection{Shake-Off}
When an electron escapes with high energy upon ionization  there is a sudden change in the potential felt by the remaining bound electrons. This may cause a subsequent ionization of another bound electron, a process referred to as shake-off. Using the sudden approximation \cite{Aberg1969, Carlson1973}, we calculate the probability for an electron to be shaken-off from the $n,l$ sub-shell as follows

\begin{equation}
\label{shakeoff}
P_{nl}\approx 1- \prod^{2l+1}_{i=1} \left[\left| \int\phi^*_{i}({H_i})\phi_{i}({H_f})d\tau \right|^2\right]^{n_i},
\end{equation}

\noindent where $\phi^*_{i}({H_i})$ and $\phi_{i}({H_f})$ are the wavefunctions for the $2l+1$ orbitals of the $n,l$ sub-shell in the initial and final Hamiltonians, respectively, and $n_i$ is the occupation of the $i$ orbital.

\section{Results}
 

Our goal is to identify the pathways leading to the formation of the charged states Xe$^{4+}$ and Xe$^{5+}$ for the pulse parameters used in the experiment described  in Ref. \cite{Bergues}. These charged states are produced when Xe interacts with a pulse of full-width half-maximum of 340 as and photon energy of 93 eV and 115 eV.  The energies needed to sequentially ionize electrons from the 4d shell are roughly equal to 70 eV for the removal of the first electron, 87 eV for the removal of the second one and 106 eV for the removal of the third electron. 

We employ a Gaussian laser pulse described in cylindrical coordinates as follows

\begin{equation}
I(r,z;t)=I(t)\dfrac{w_0^2}{w(z)^2}exp\Big[ {\dfrac{-2r^2}{w(z)^2}} \Big],
\label{eq:GaussianPulse}
\end{equation}

\noindent where $r$ is the radius and $z$ is the beam propagation axis.  The beam waist is denoted by $w_0$, which is equal to 0.85 $\mu$m for the 93 eV pulse and 2.12 $\mu$m for the 115 eV pulse. The beam radius at a distance $z$ is given below

\begin{equation}
w(z)=w_0\sqrt{1+(z/z_R)^2},
\label{wz}
\end{equation}

\noindent where $z_R$ is the Rayleigh length and is equal to 93 $\mu$m for both pulses. Furthermore, to calculate the ion yields and the prevalent pathways, we perform a volume averaging. To do so we consider a grid $(r,z)$ consisting of equidistant points. Namely, r varies from 0 $\mu$m to 4.82 $\mu$m in steps 
of 0.01 $\mu$m and z varies from -10$^4$ $\mu$m to 10$^4$ $\mu$m in steps of 0.5 $\mu$m. These grid points were chosen so that we obtain good convergence for the ion yields. At each grid point we compute the intensity of the pulse in accordance with Eq. (\ref{eq:GaussianPulse}). The ion yields are then calculated for each grid point. The sum of the respective yields of all grid points give us the total yield for each ion state. 
\newline

\subsection{Ion yields and ratios of ion yields}
In what follows, we first compare our results with experimental ones for relative ion yields \cite{Bergues, Holland_1979} and ion yield ratios  \cite{Bergues}. In Ref. \cite{Bergues}, the experimental pulse was obtained by high-harmonic generation while Ref. \cite{Holland_1979} involves synchrotron radiation. To account for the uncertainty in the intensity of the experimental results, we consider intensities equal to $10^{14}$ $Wcm^{-2}$, 8$\times10^{13}$ $Wcm^{-2}$ and 6$\times10^{13}$ $Wcm^{-2}$. Tables \ref{tab:IonYields_93} and \ref{tab:IonYields115} show that our results for ion charges up to Xe$^{3+}$ are in reasonable agreement with the experimental results  for the  charged states Xe$^{2+}$ and Xe$^{3+}$. \newline

\begin{table}
\caption{\label{tab:IonYields_93} Relative ion yields and yield rations for Xe interacting with an XUV pulse of photon energy 93 eV. The intensity is given in units of $Wcm^{-2}$. The yields of all charged states add up to 100.  } \begin{ruledtabular}.
\begin{tabular}{llllll}
Ion & Ref. \protect\cite{Bergues} & Ref. \protect\cite{Holland_1979} & \multicolumn{3}{c}{\pbox{5cm}{This work}} \\
\hline
 & & & $10^{14}$ & 8$\times10^{13}$  & 6$\times10^{13}$  \\
\hline
Xe$^{+}$ & 3.4 & 5.7 & 1.42 & 1.42 & 1.42 \\
Xe$^{2+}$ & 77.6 & 68.6 & 74.8 & 74.8 & 74.8\\ 
Xe$^{3+}$ & 19.0 & 25.7 & 23.7& 23.7 & 23.7 \\
$ \dfrac{Xe^{4+}}{Xe^{2+}} $ & 4.0$\times10^{-3}$ & - & 1.5$\times10^{-2}$ & 1.2$\times10^{-2}$ & 8.9$\times10^{-3}$ \\
\end{tabular}
\end{ruledtabular}
\end{table}

\begin{center}
\begin{figure}[h!]
\includegraphics[width=0.3\textwidth ]{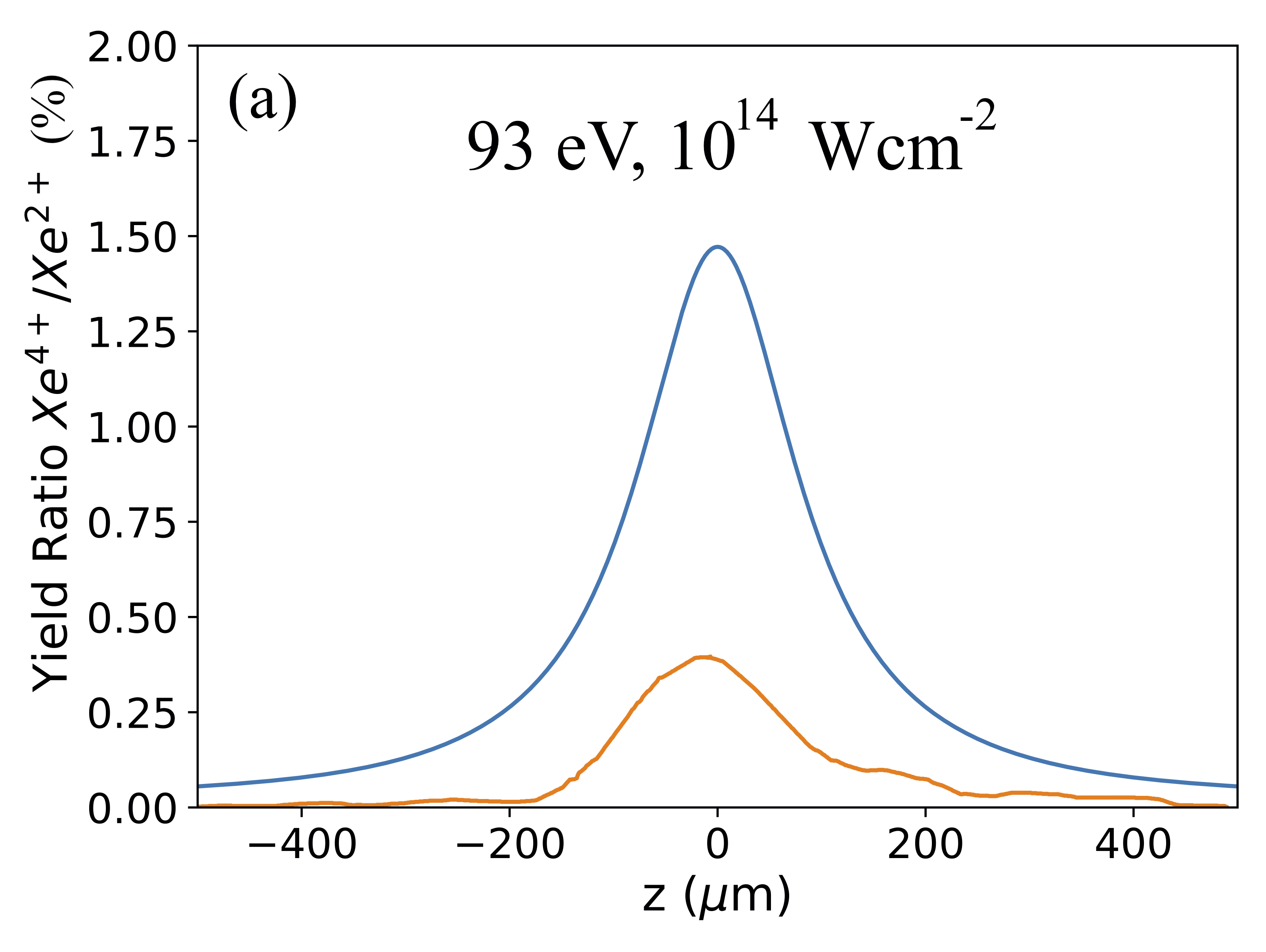}
\includegraphics[width=0.3\textwidth]{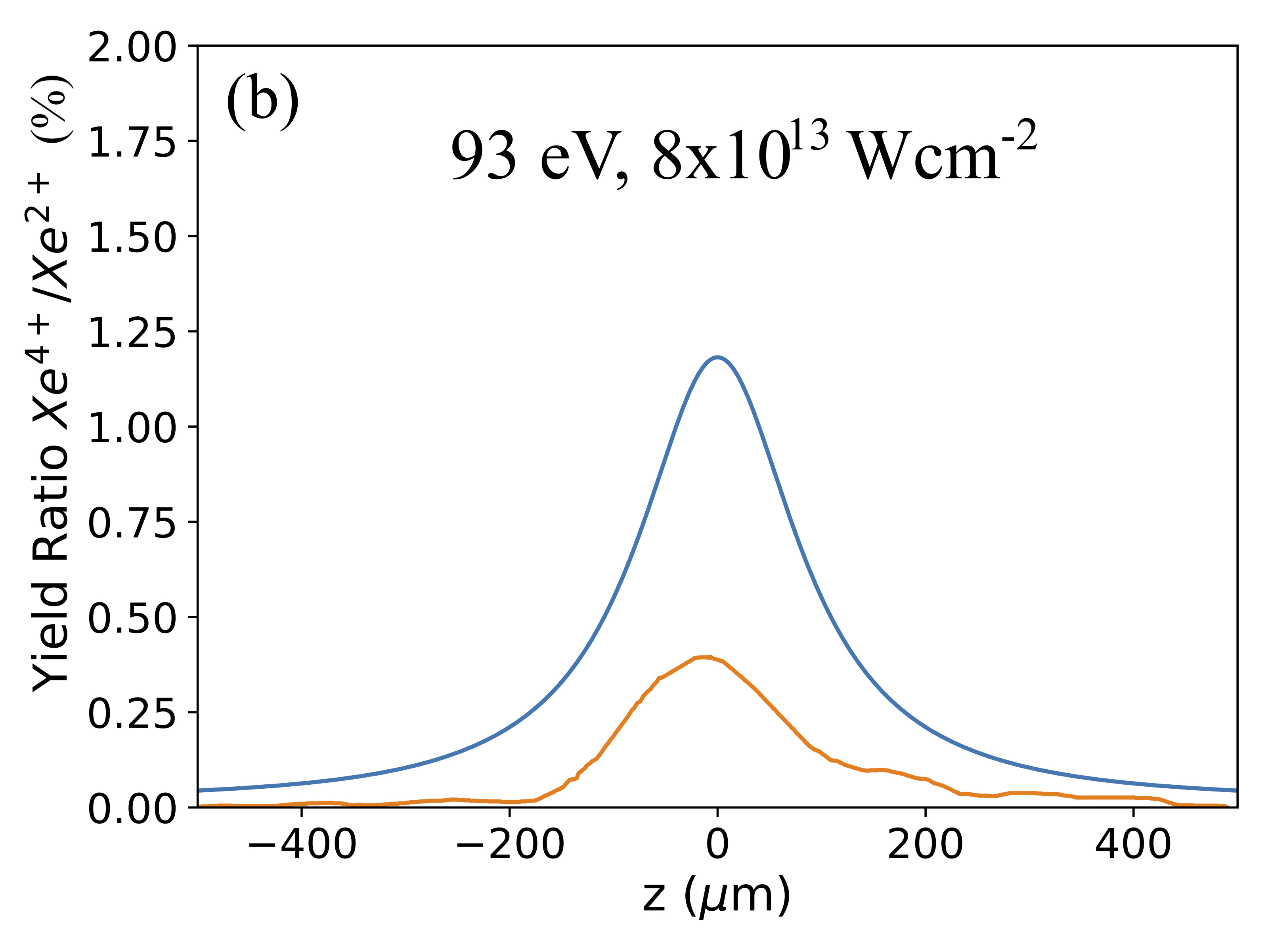}
\includegraphics[width=0.3\textwidth]{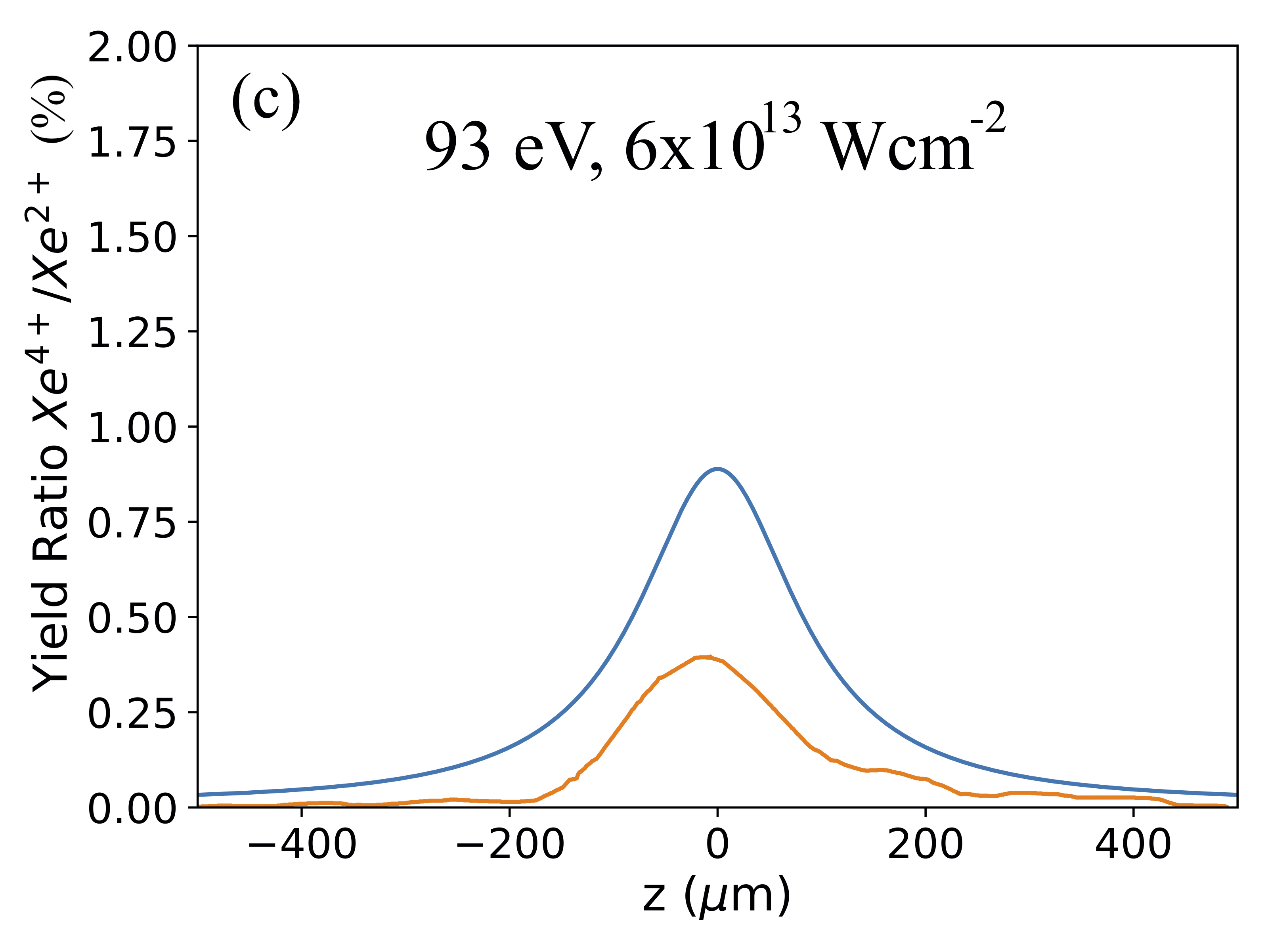}
\includegraphics[width=0.3\textwidth ]{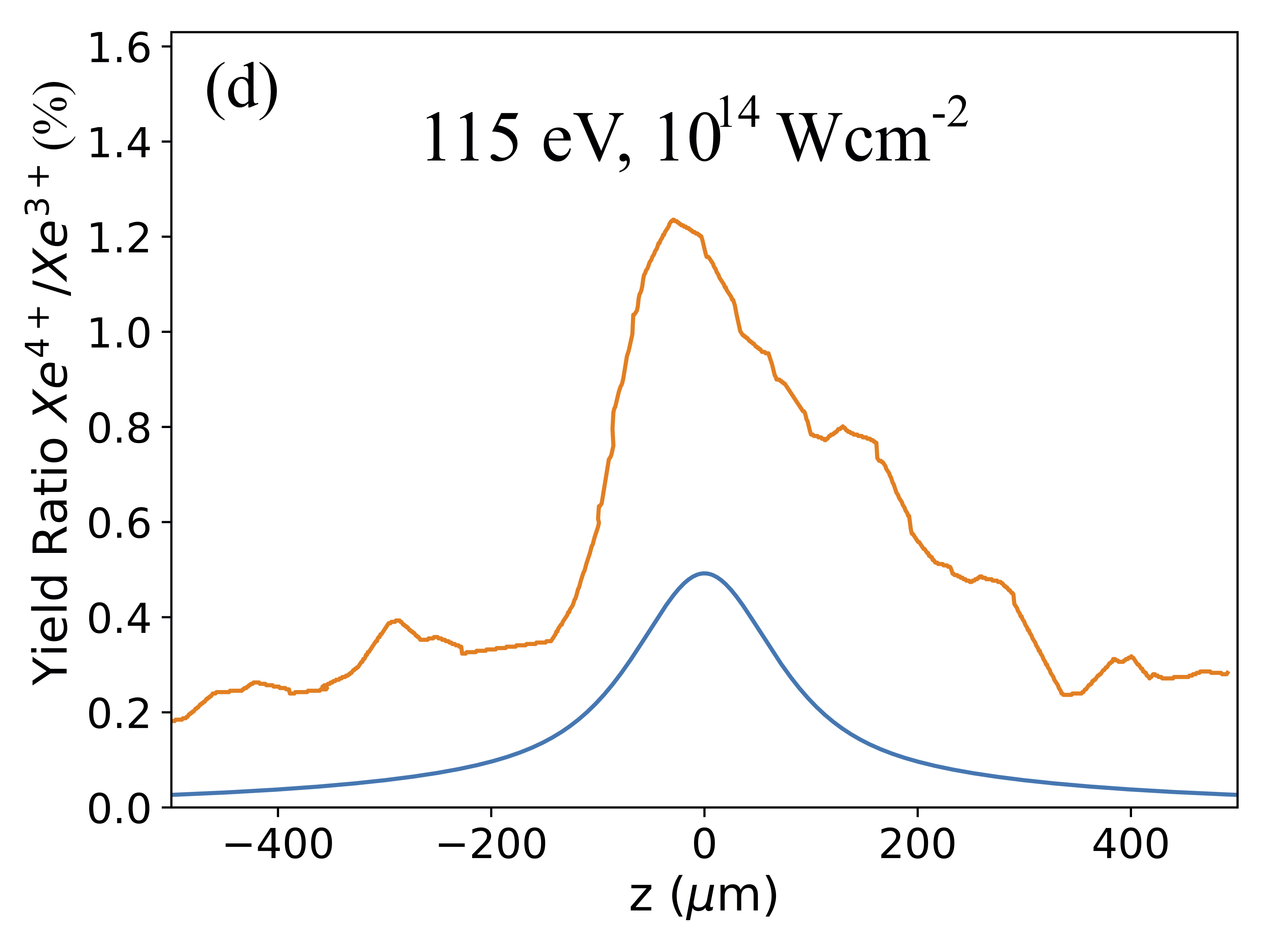}
\includegraphics[width=0.3\textwidth]{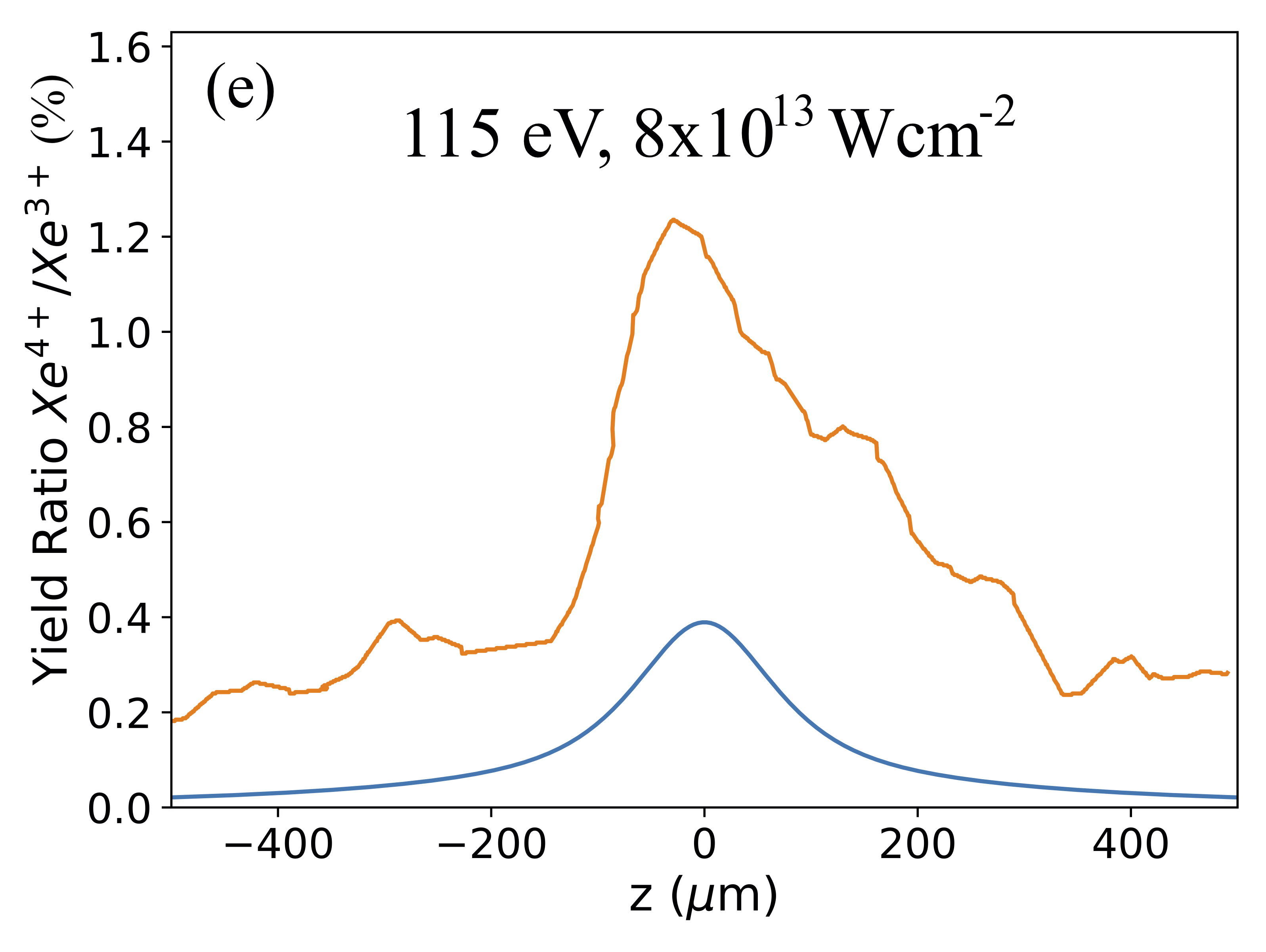}
\includegraphics[width=0.3\textwidth]{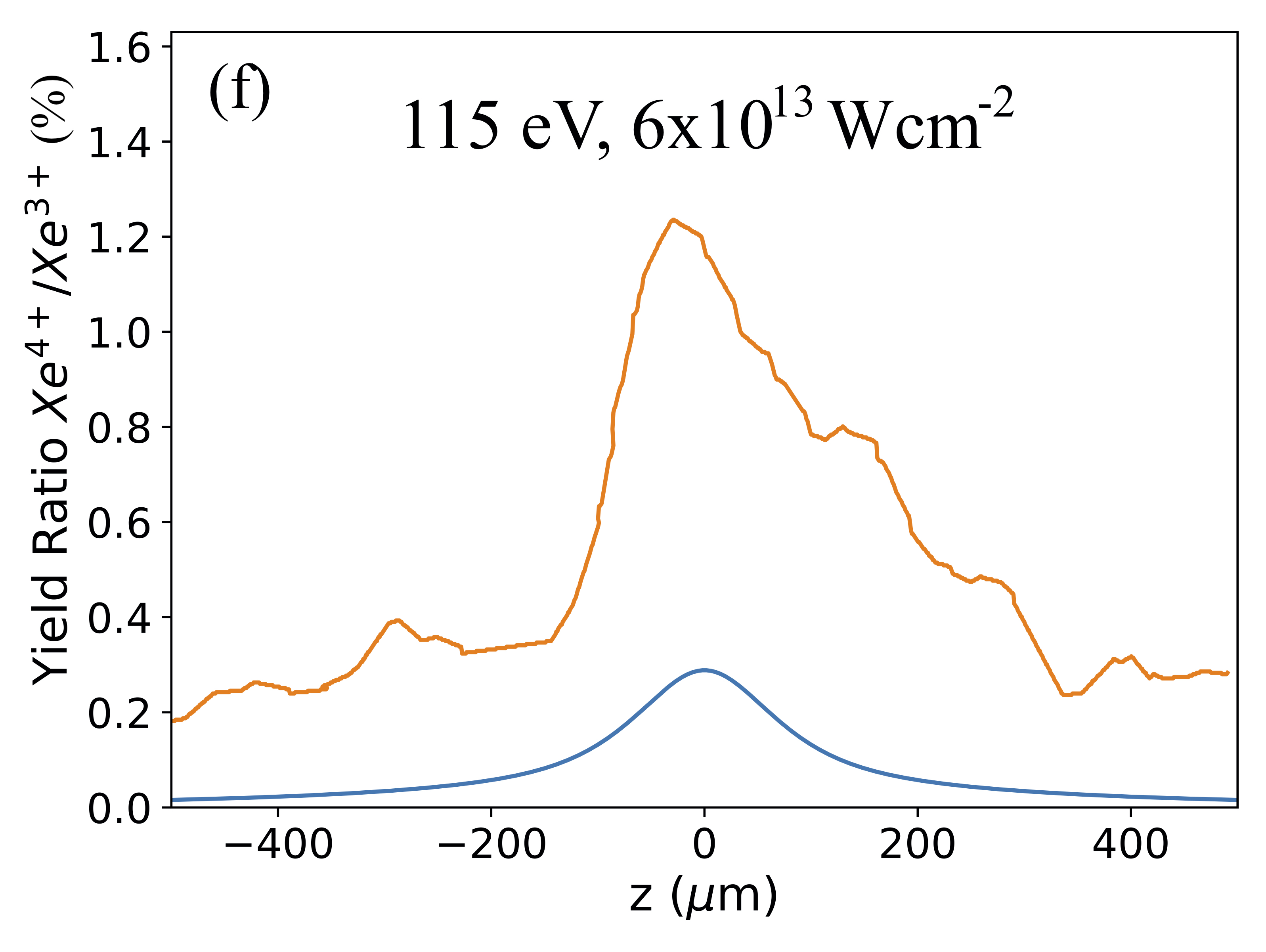}
\caption{\label{YieldRatios_Fig}Yield ratio as a function of the  propagation distance in the focus. Orange lines denote the experimental results adapted from Ref. \protect\cite{Bergues} and blue lines denote our own ratios of the ion yields. }
\end{figure}
\end{center}

In Table \ref{tab:IonYields_93}, we also show the ratio of the Xe$^{4+}$ and Xe$^{2+}$  ion yields  for the 93 eV pulse. We find that the difference with the experimental ratio in Ref. \cite{Bergues} depends on the intensity considered and roughly amounts to a factor of two for 6$\times10^{13}$  $Wcm^{-2}$. Moreover, in Table III we compare the ratio of the ion yields Xe$^{4+}$ and Xe$^{3+}$ with the experimental ratio \cite{Bergues} for the 115 eV pulse. We find that the ratio we  compute differs by roughly a factor of two  from the experimental result for $10^{14}$  $Wcm^{-2}$.  We note that the ion yields for Xe$^{4+}$ and Xe$^{5+}$ are subjected to an experimental statistical uncertainty of up to 15 \%. 
The deviations between the computed and the experimental values for the  above ratios of the ion yields may be also partially explained by the experimental uncertainty in the pulse duration and intensity.  In addition, in Fig. \ref{YieldRatios_Fig} we plot the dependence on the propagation axis $z$ of the ratio Xe$^{4+}$/Xe$^{2+}$ for the 93 eV pulse and of the ratio Xe$^{4+}$/Xe$^{3+}$ for the 115 eV pulse, for three different intensities. We believe that the agreement between theory and experiment within a factor of 2 is reasonable in view of  the experimental uncertainties.

\begin{table}
\caption{\label{tab:IonYields115}Relative ion yields and yield rations for Xe interacting with an XUV pulse of photon energy 115 eV. The intensity is given in units of $Wcm^{-2}$. The yields of all charged states add up to 100.} \begin{ruledtabular}
\begin{tabular}{llllll}
Ion & Ref. \protect\cite{Bergues} & Ref. \protect\cite{Holland_1979} & \multicolumn{3}{c}{\pbox{5cm}{This work}} \\
\hline
 & & & $10^{14}$ & 8$\times10^{13}$  & 6$\times10^{13}$  \\
\hline
Xe$^{+}$ & - & 2.95 & 7.23 & 7.23 & 7.23 \\
Xe$^{2+}$ & - & 69.2 & 70.5& 70.5 & 70.5 \\ 
Xe$^{3+}$ & - & 27.8 & 22.3 & 22.3 & 22.3 \\
$\dfrac{Xe^{4+}}{Xe^{3+}}$ & 1.2$x10^{-2}$  & - & 4.9$\times10^{-3}$ & 3.9$\times10^{-3}$ & 2.9$\times10^{-3}$ \\
\end{tabular}
\end{ruledtabular}
\end{table}



\onecolumngrid
\twocolumngrid

 
 \subsection{Pathways}
\begin{figure}[h!]
\includegraphics[width=\linewidth]{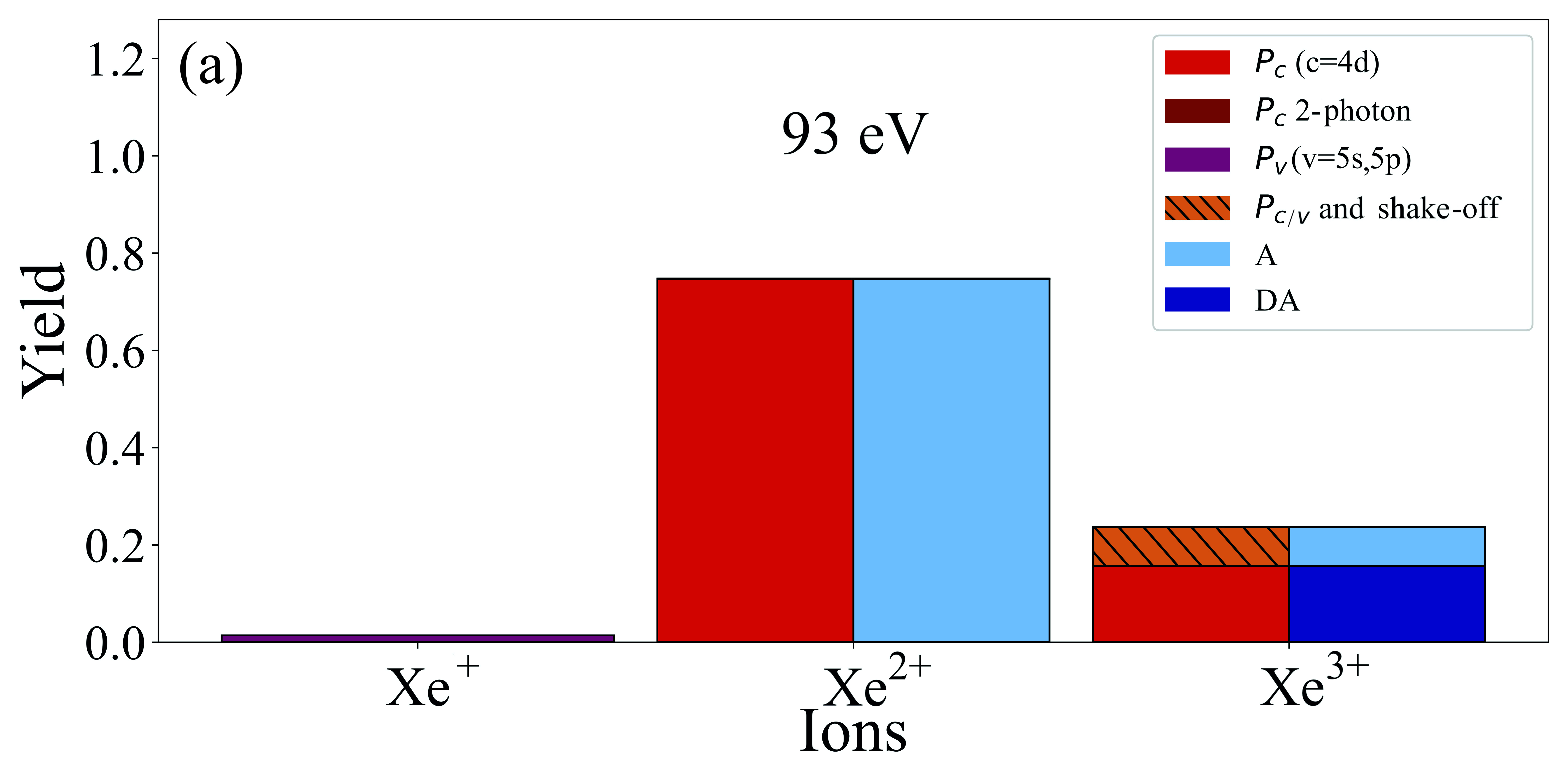}
\includegraphics[width=\linewidth]{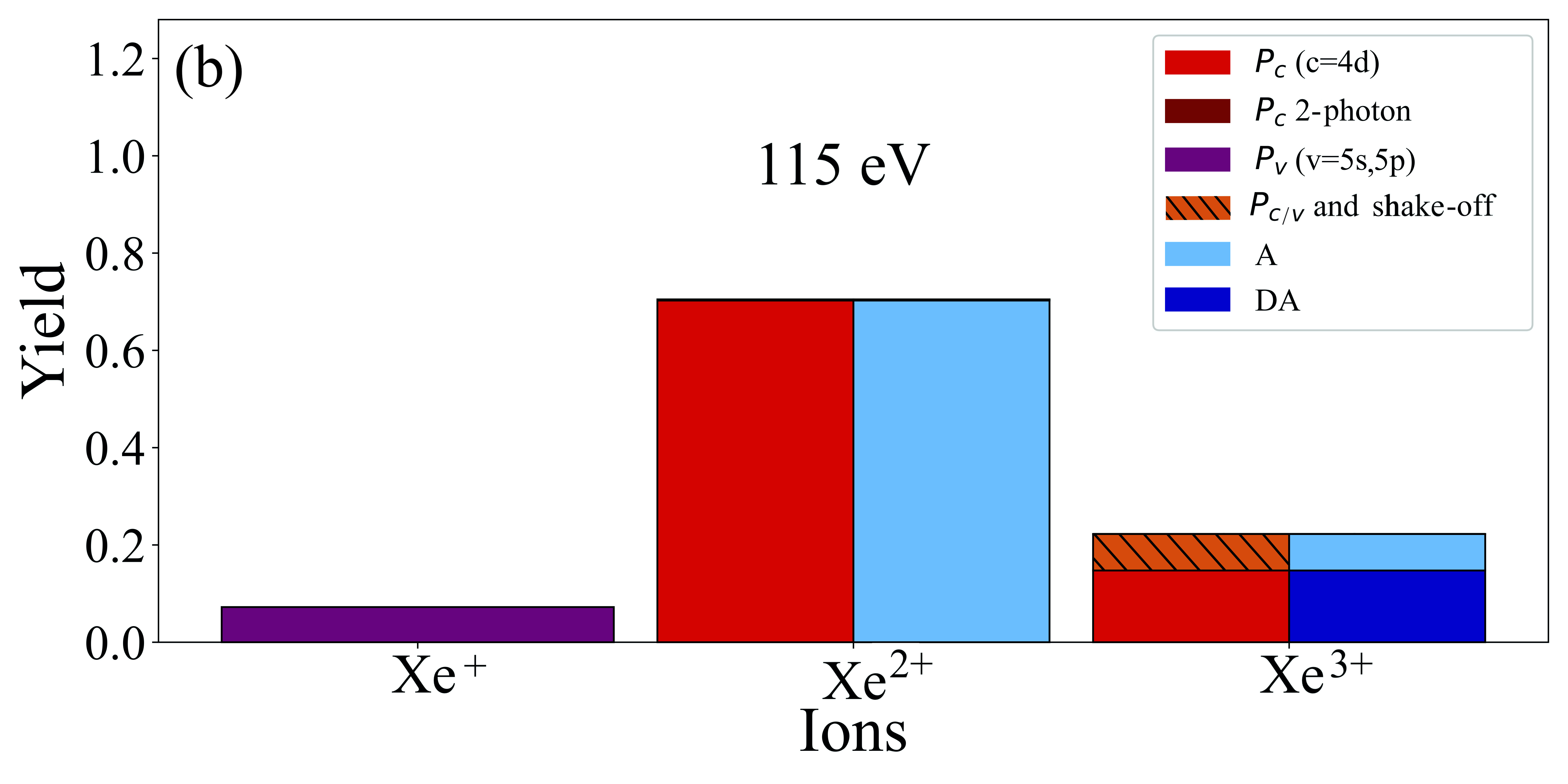}
\caption{Pathways leading to the formation of ion charges Xe$^+$, Xe$^{2+}$ and Xe$^{3+} $ for two different photon energies of 93 eV and 115 eV. The vertical axis shows the yield of each ion state on the horizontal axis. The yields of all charged states add up to 1. Each column corresponds to a different process. For each charged state, the sequence of the processes takes place from left to right. $P_c$ $(c=4d)$ stands for ionization of a $4d$ electron by single-photon absorption; $P_c$ 2-photon stands for ionization of a $4d$ electron by two-photon absorption; $P_v$ ($v=5s,5p$) stands for ionization of a valence electron $5s$ or $5p$ by single-photon absorption; $P_{c/v}$ and shake-off stands for ionization of a core or valence electron by single-photon absorption followed by ionization of another electron due to shake-off; A and DA stand for Auger decay and double Auger decay, respectively. The intensity considered is $10^{14}$  $Wcm^{-2}$. }
\label{Pathways_1_2_3}
\end{figure}

Next, we identify the prevalent pathways that lead to the formation of Xe ion states Xe$^{n+}$, where $n=1,2,3$. These pathways are shown in Fig. \ref{Pathways_1_2_3}(a) for the 93 eV pulse and in Fig. \ref{Pathways_1_2_3}(b) for the 115 eV pulse. In Fig. \ref{Pathways_1_2_3} the vertical axis corresponds to the relative ion yield of each ion state, where the latter is shown on the horizontal axis. The sum of the yields of all charged states of Xe$^{n+}$, with $n=1-5$, is equal to 1. Figs. \ref{Pathways_1_2_3}(a)-(b) show that the prevalent pathway leading to the formation of Xe$^{+}$ is ionization of a valence electron by single-photon absorption ($P_v$ $(v=5s,5p)$). We also find that Xe$^{2+}$ is formed by a sequence of two processes. The first process involves ionization of a core electron by single-photon absorption ($P_c$ $(c=4d)$). The subsequent process is a single Auger decay ($A$). In addition, we find that Xe$^{3+}$ is formed mainly by ionization of a core electron by single-photon absorption ($P_c$ $(c=4d)$) followed by a double Auger process ($DA$), i.e. an electron fills in the 4d core hole, while two other electrons escape. Hence, Xe$^{3+}$ is formed by a sequence of a single-photon absorption process and a double Auger one. 
As expected, our results for the prevalent pathways leading to the formation of Xe$^{+}$, Xe$^{2+}$ and Xe$^{3+}$ are consistent with a slope equal to one on a log-log scale of the ion yields as a function of intensity  \cite{Holland_1979},  see Fig. 3.

\begin{figure}[ht]
\includegraphics[width=\linewidth]{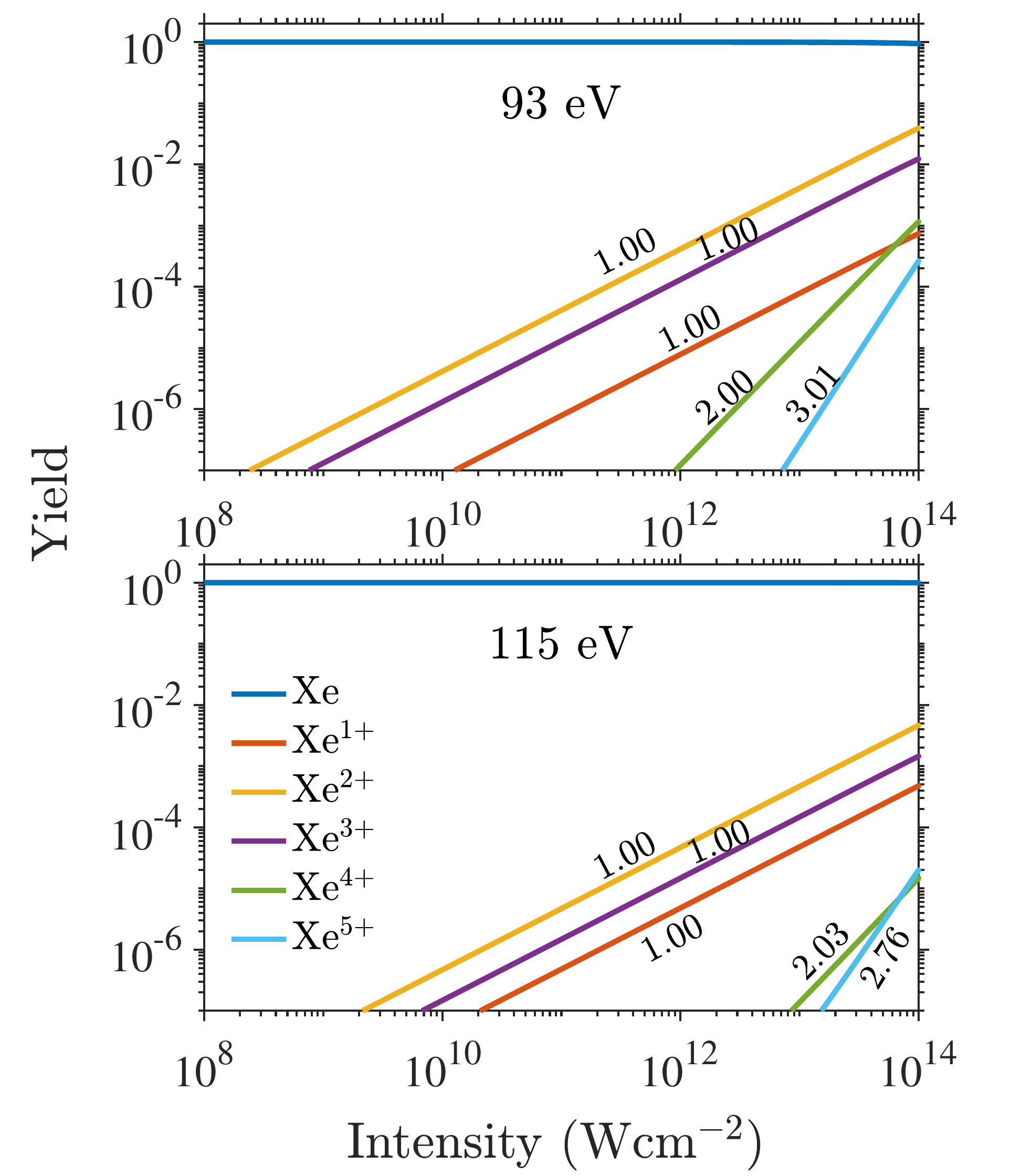}
\caption{Log-log plot of the ion yields versus pulse intensity for the 93 eV pulse (a) and for the 115 eV pulse (b). The numbers along the lines correspond to the slope of each of the yields versus intensity.}
\label{ionYields_Fig}
\end{figure}

\begin{center}
\begin{figure}[h!]
\includegraphics[scale=0.06]{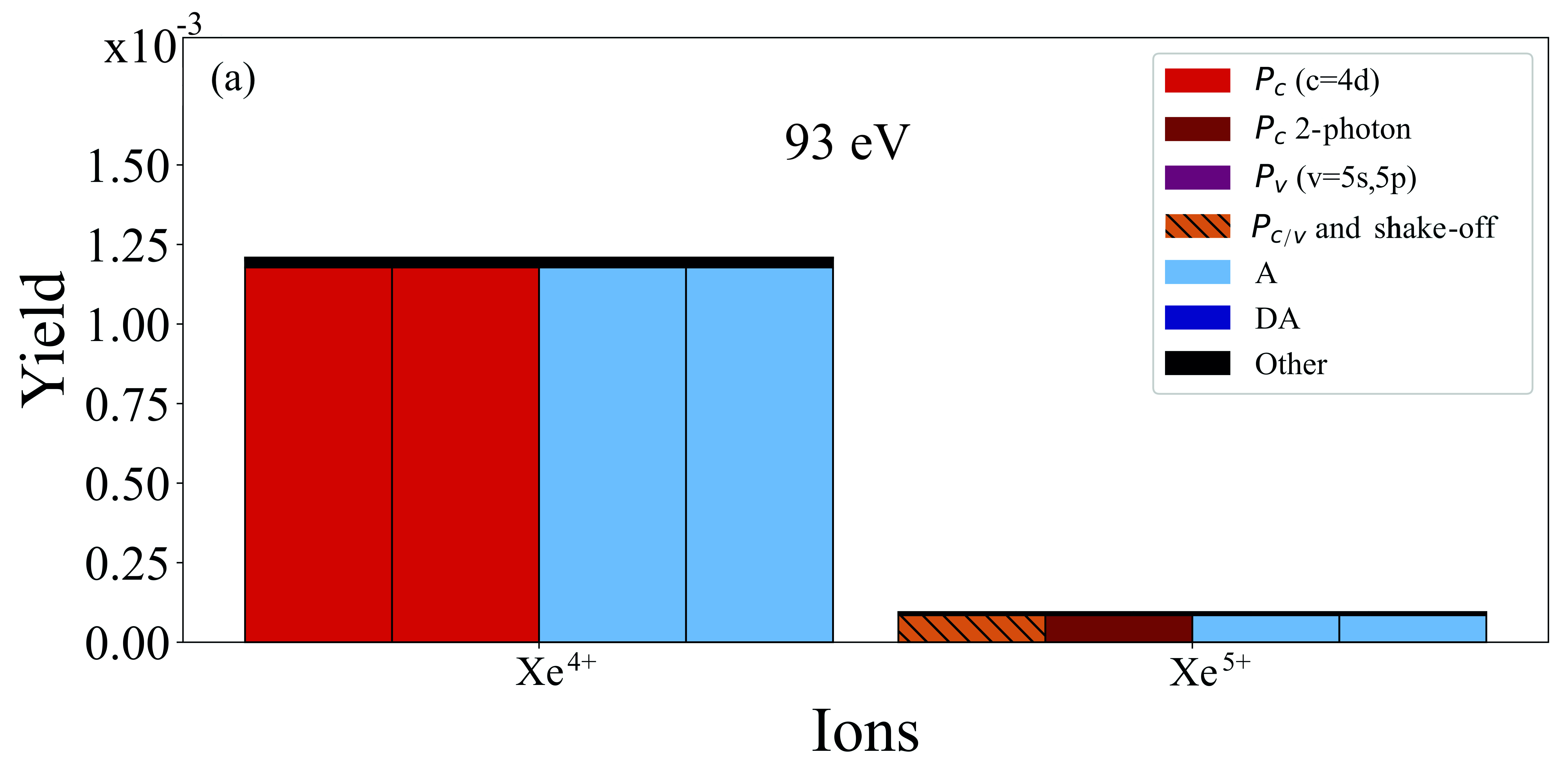}
\includegraphics[scale=0.06]{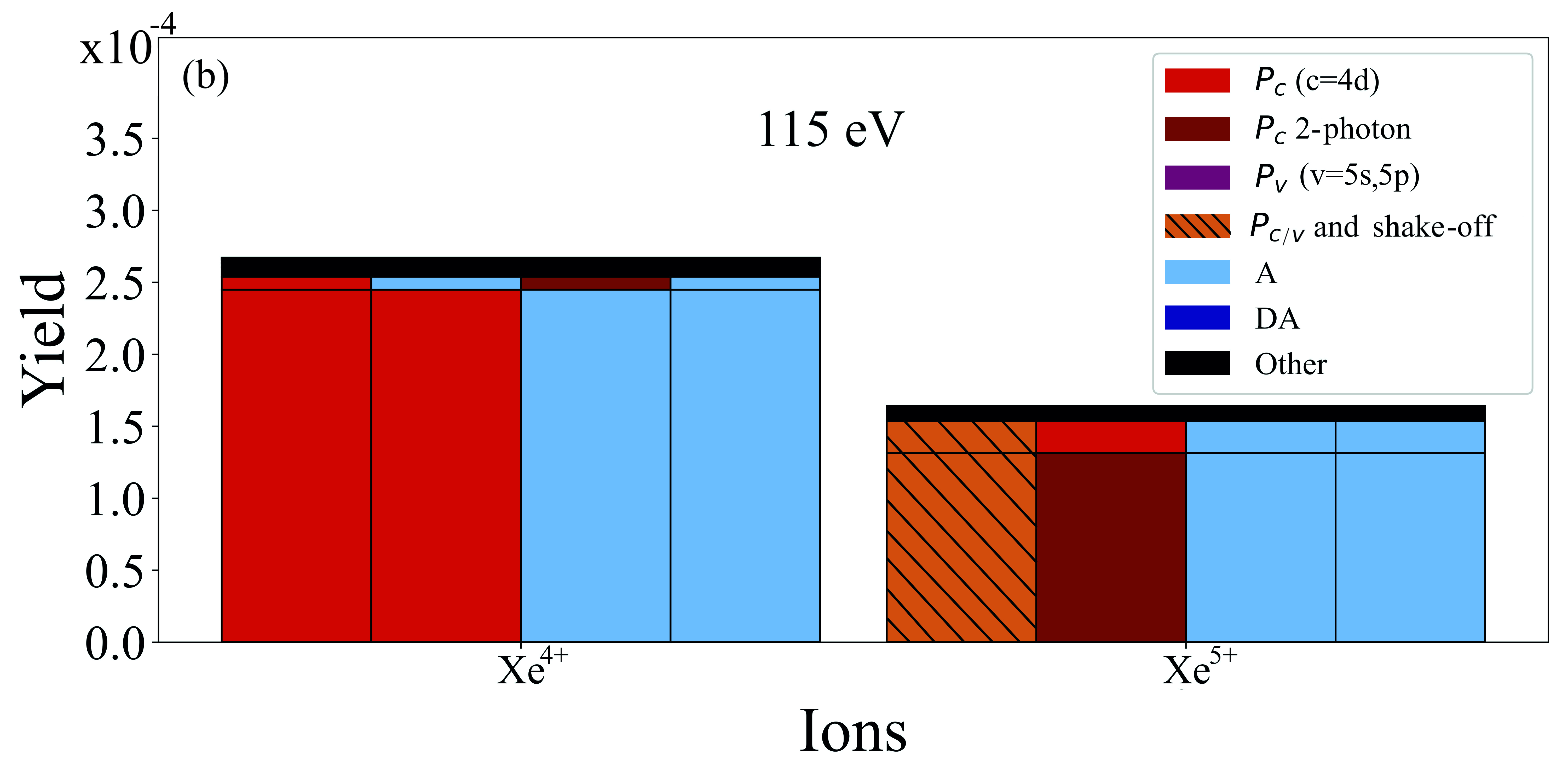}
\caption{\label{Pathways_4_5}  Same as Fig. \protect\ref{Pathways_1_2_3} for charged states Xe$^{4+}$ and Xe$^{5+}$.}
\end{figure}
\end{center}

In Fig. \ref{Pathways_4_5}(a) for the 93 eV pulse and Fig. \ref{Pathways_4_5}(b) for the 115 eV pulse, we show the prevalent pathways for charged states Xe$^{4+}$ and Xe$^{5+}$. We find that the prevalent pathway leading to the formation of Xe$^{4+}$ consists of a sequence of four processes. First, a $4d$ core electron is ionized by a single-photon absorption  ($P_c$ $(c=4d)$). Then, before the Xe$^{+}$ ion relaxes, another $4d$ core electron is ionized via single-photon absorption. Thus, the first two electrons are ionized by two sequential single-photon absorption processes forming a double core-hole state. This is a process that was not accounted for in Ref. \cite{Bergues}. The third and fourth electrons are ionized by a sequence of two single Auger processes. Therefore, we find that the prevalent pathway leading to the formation of Xe$^{4+}$ involves the absorption of two photons. This is consistent with the slope of the yield versus intensity of Xe$^{4+}$ being equal to two, see Fig. \ref{ionYields_Fig}.

For both the 93 eV and the 115 eV pulses, we find that Xe$^{5+}$ is formed mainly by one pathway that involves four processes. The first two electrons are ionized by a single-photon absorption followed by shake-off  ($P_{c/v}$ and shake-off). Next, a direct two-photon ionization process takes place ($P_c$ $2-photon$). That is, a $4d$ core electron escapes by absorbing two photons. Following the two ionization processes, two Auger decays take place, one after the other, resulting in the emission of the fourth and the fifth electron.  It is quite interesting that Xe$^{5+}$ is formed by a pathway involving a two-photon ionization process when Xe interacts with an attosecond XUV pulse. In previous studies of Xe interacting with a femtosecond XUV pulse, two-photon ionization processes were found to play a significant role only for ion states higher than Xe$^{7+}$ \cite{Lambropoulos_21}.  Energetically, two photons would suffice for the formation of Xe$^{5+}$. Surprisingly, we find that Xe$^{5+}$ is preferentially created via absorption of  three photons. As expected, this is reflected in  the slope being roughly equal to three for Xe$^{5+}$  in Fig. \ref{ionYields_Fig}.

\onecolumngrid
\twocolumngrid

\section{Conclusion}
We have identified the main pathways leading to the formation of charged states up to Xe$^{5+}$ when it interacts with an attosecond XUV pulse. Both for Xe$^{4+}$ and for Xe$^{5+}$ we find that the main pathway for their formation proceeds via two sequential photo-absorption processes, i.e. via the formation of a double core-hole state. For Xe$^{4+}$ these sequential photo-ionization processes involve each one photon. However, for Xe$^{5+}$ one of the two sequential photo-ionization processes involves a direct two-photon absorption process. So far, such direct two-photon absorption was only identified for the formation of charged states 
higher than Xe$^{7+}$ for interaction with femtosecond XUV pulses \cite{Lambropoulos_21}.



\acknowledgements
We acknowledge the use of the Legion computational resources at UCL. This work was funded by the Leverhulme Trust Research Project Grant 2017-376. B. B acknowledges fruitful discussions with Hartmut Schr$\ddot{o}$der and is grateful for support from Matthias Kling.

\newpage

\bibliography{XePaper}{}

\begin{thebibliography}{39}%
\makeatletter
\providecommand \@ifxundefined [1]{%
 \@ifx{#1\undefined}
}%
\providecommand \@ifnum [1]{%
 \ifnum #1\expandafter \@firstoftwo
 \else \expandafter \@secondoftwo
 \fi
}%
\providecommand \@ifx [1]{%
 \ifx #1\expandafter \@firstoftwo
 \else \expandafter \@secondoftwo
 \fi
}%
\providecommand \natexlab [1]{#1}%
\providecommand \enquote  [1]{``#1''}%
\providecommand \bibnamefont  [1]{#1}%
\providecommand \bibfnamefont [1]{#1}%
\providecommand \citenamefont [1]{#1}%
\providecommand \href@noop [0]{\@secondoftwo}%
\providecommand \href [0]{\begingroup \@sanitize@url \@href}%
\providecommand \@href[1]{\@@startlink{#1}\@@href}%
\providecommand \@@href[1]{\endgroup#1\@@endlink}%
\providecommand \@sanitize@url [0]{\catcode `\\12\catcode `\$12\catcode
  `\&12\catcode `\#12\catcode `\^12\catcode `\_12\catcode `\%12\relax}%
\providecommand \@@startlink[1]{}%
\providecommand \@@endlink[0]{}%
\providecommand \url  [0]{\begingroup\@sanitize@url \@url }%
\providecommand \@url [1]{\endgroup\@href {#1}{\urlprefix }}%
\providecommand \urlprefix  [0]{URL }%
\providecommand \Eprint [0]{\href }%
\providecommand \doibase [0]{http://dx.doi.org/}%
\providecommand \selectlanguage [0]{\@gobble}%
\providecommand \bibinfo  [0]{\@secondoftwo}%
\providecommand \bibfield  [0]{\@secondoftwo}%
\providecommand \translation [1]{[#1]}%
\providecommand \BibitemOpen [0]{}%
\providecommand \bibitemStop [0]{}%
\providecommand \bibitemNoStop [0]{.\EOS\space}%
\providecommand \EOS [0]{\spacefactor3000\relax}%
\providecommand \BibitemShut  [1]{\csname bibitem#1\endcsname}%
\let\auto@bib@innerbib\@empty
\bibitem [{\citenamefont {Pellegrini}(2012)}]{Pellegrini2012}%
  \BibitemOpen
  \bibfield  {author} {\bibinfo {author} {\bibfnamefont {C.}~\bibnamefont
  {Pellegrini}},\ }\href {\doibase 10.1140/epjh/e2012-20064-5} {\bibfield
  {journal} {\bibinfo  {journal} {The European Physical Journal H}\ }\textbf
  {\bibinfo {volume} {37}},\ \bibinfo {pages} {659} (\bibinfo {year}
  {2012})}\BibitemShut {NoStop}%
\bibitem [{\citenamefont {Marangos}(2011)}]{Marangos_2011}%
  \BibitemOpen
  \bibfield  {author} {\bibinfo {author} {\bibfnamefont {J.}~\bibnamefont
  {Marangos}},\ }\href@noop {} {\bibfield  {journal} {\bibinfo  {journal}
  {Contemporary Physics}\ }\textbf {\bibinfo {volume} {52}},\ \bibinfo {pages}
  {551} (\bibinfo {year} {2011})}\BibitemShut {NoStop}%
\bibitem [{\citenamefont {Ullrich}\ \emph {et~al.}(2012)\citenamefont
  {Ullrich}, \citenamefont {Rudenko},\ and\ \citenamefont
  {Moshammer}}]{Ullrich_2012}%
  \BibitemOpen
  \bibfield  {author} {\bibinfo {author} {\bibfnamefont {J.}~\bibnamefont
  {Ullrich}}, \bibinfo {author} {\bibfnamefont {A.}~\bibnamefont {Rudenko}}, \
  and\ \bibinfo {author} {\bibfnamefont {R.}~\bibnamefont {Moshammer}},\
  }\href@noop {} {\bibfield  {journal} {\bibinfo  {journal} {Annual review of
  physical chemistry}\ }\textbf {\bibinfo {volume} {63}},\ \bibinfo {pages}
  {635} (\bibinfo {year} {2012})}\BibitemShut {NoStop}%
\bibitem [{\citenamefont {Wallis}\ \emph {et~al.}(2015)\citenamefont {Wallis},
  \citenamefont {Banks},\ and\ \citenamefont {Emmanouilidou}}]{Wallis2015}%
  \BibitemOpen
  \bibfield  {author} {\bibinfo {author} {\bibfnamefont {A.~O.~G.}\
  \bibnamefont {Wallis}}, \bibinfo {author} {\bibfnamefont {H.~I.~B.}\
  \bibnamefont {Banks}}, \ and\ \bibinfo {author} {\bibfnamefont
  {A.}~\bibnamefont {Emmanouilidou}},\ }\href {\doibase
  10.1103/PhysRevA.91.063402} {\bibfield  {journal} {\bibinfo  {journal} {Phys.
  Rev. A}\ }\textbf {\bibinfo {volume} {91}},\ \bibinfo {pages} {063402}
  (\bibinfo {year} {2015})}\BibitemShut {NoStop}%
\bibitem [{\citenamefont {Pi}\ and\ \citenamefont
  {Starace}(2010)}]{Pi_XeStudy}%
  \BibitemOpen
  \bibfield  {author} {\bibinfo {author} {\bibfnamefont {L.-W.}\ \bibnamefont
  {Pi}}\ and\ \bibinfo {author} {\bibfnamefont {A.~F.}\ \bibnamefont
  {Starace}},\ }\href {\doibase 10.1103/PhysRevA.82.053414} {\bibfield
  {journal} {\bibinfo  {journal} {Phys. Rev. A}\ }\textbf {\bibinfo {volume}
  {82}},\ \bibinfo {pages} {053414} (\bibinfo {year} {2010})}\BibitemShut
  {NoStop}%
\bibitem [{\citenamefont {Richardson}\ \emph {et~al.}(2010)\citenamefont
  {Richardson}, \citenamefont {Costello}, \citenamefont {Cubaynes},
  \citenamefont {D\"usterer}, \citenamefont {Feldhaus}, \citenamefont {van~der
  Hart}, \citenamefont {Jurani\ifmmode~\acute{c}\else \'{c}\fi{}},
  \citenamefont {Li}, \citenamefont {Meyer}, \citenamefont {Richter},
  \citenamefont {Sorokin},\ and\ \citenamefont {Tiedke}}]{Costello_XeStudy}%
  \BibitemOpen
  \bibfield  {author} {\bibinfo {author} {\bibfnamefont {V.}~\bibnamefont
  {Richardson}}, \bibinfo {author} {\bibfnamefont {J.~T.}\ \bibnamefont
  {Costello}}, \bibinfo {author} {\bibfnamefont {D.}~\bibnamefont {Cubaynes}},
  \bibinfo {author} {\bibfnamefont {S.}~\bibnamefont {D\"usterer}}, \bibinfo
  {author} {\bibfnamefont {J.}~\bibnamefont {Feldhaus}}, \bibinfo {author}
  {\bibfnamefont {H.~W.}\ \bibnamefont {van~der Hart}}, \bibinfo {author}
  {\bibfnamefont {P.}~\bibnamefont {Jurani\ifmmode~\acute{c}\else \'{c}\fi{}}},
  \bibinfo {author} {\bibfnamefont {W.~B.}\ \bibnamefont {Li}}, \bibinfo
  {author} {\bibfnamefont {M.}~\bibnamefont {Meyer}}, \bibinfo {author}
  {\bibfnamefont {M.}~\bibnamefont {Richter}}, \bibinfo {author} {\bibfnamefont
  {A.~A.}\ \bibnamefont {Sorokin}}, \ and\ \bibinfo {author} {\bibfnamefont
  {K.}~\bibnamefont {Tiedke}},\ }\href {\doibase
  10.1103/PhysRevLett.105.013001} {\bibfield  {journal} {\bibinfo  {journal}
  {Phys. Rev. Lett.}\ }\textbf {\bibinfo {volume} {105}},\ \bibinfo {pages}
  {013001} (\bibinfo {year} {2010})}\BibitemShut {NoStop}%
\bibitem [{\citenamefont {Richter}\ \emph {et~al.}(2009)\citenamefont
  {Richter}, \citenamefont {Amusia}, \citenamefont {Bobashev}, \citenamefont
  {Feigl}, \citenamefont {Jurani\ifmmode~\acute{c}\else \'{c}\fi{}},
  \citenamefont {Martins}, \citenamefont {Sorokin},\ and\ \citenamefont
  {Tiedtke}}]{Amusia_XeStudy}%
  \BibitemOpen
  \bibfield  {author} {\bibinfo {author} {\bibfnamefont {M.}~\bibnamefont
  {Richter}}, \bibinfo {author} {\bibfnamefont {M.~Y.}\ \bibnamefont {Amusia}},
  \bibinfo {author} {\bibfnamefont {S.~V.}\ \bibnamefont {Bobashev}}, \bibinfo
  {author} {\bibfnamefont {T.}~\bibnamefont {Feigl}}, \bibinfo {author}
  {\bibfnamefont {P.~N.}\ \bibnamefont {Jurani\ifmmode~\acute{c}\else
  \'{c}\fi{}}}, \bibinfo {author} {\bibfnamefont {M.}~\bibnamefont {Martins}},
  \bibinfo {author} {\bibfnamefont {A.~A.}\ \bibnamefont {Sorokin}}, \ and\
  \bibinfo {author} {\bibfnamefont {K.}~\bibnamefont {Tiedtke}},\ }\href
  {\doibase 10.1103/PhysRevLett.102.163002} {\bibfield  {journal} {\bibinfo
  {journal} {Phys. Rev. Lett.}\ }\textbf {\bibinfo {volume} {102}},\ \bibinfo
  {pages} {163002} (\bibinfo {year} {2009})}\BibitemShut {NoStop}%
\bibitem [{\citenamefont {Son}\ \emph {et~al.}(2012)\citenamefont {Son},
  \citenamefont {Santra} \emph {et~al.}}]{Son_2012}%
  \BibitemOpen
  \bibfield  {author} {\bibinfo {author} {\bibfnamefont {S.-K.}\ \bibnamefont
  {Son}}, \bibinfo {author} {\bibfnamefont {R.}~\bibnamefont {Santra}},  \emph
  {et~al.},\ }\href@noop {} {\bibfield  {journal} {\bibinfo  {journal}
  {Physical Review A}\ }\textbf {\bibinfo {volume} {85}},\ \bibinfo {pages}
  {063415} (\bibinfo {year} {2012})}\BibitemShut {NoStop}%
\bibitem [{\citenamefont {Toyota}\ \emph {et~al.}(2017)\citenamefont {Toyota},
  \citenamefont {Son}, \citenamefont {Santra} \emph {et~al.}}]{Toyota_2017}%
  \BibitemOpen
  \bibfield  {author} {\bibinfo {author} {\bibfnamefont {K.}~\bibnamefont
  {Toyota}}, \bibinfo {author} {\bibfnamefont {S.-K.}\ \bibnamefont {Son}},
  \bibinfo {author} {\bibfnamefont {R.}~\bibnamefont {Santra}},  \emph
  {et~al.},\ }\href@noop {} {\bibfield  {journal} {\bibinfo  {journal}
  {Physical Review A}\ }\textbf {\bibinfo {volume} {95}},\ \bibinfo {pages}
  {043412} (\bibinfo {year} {2017})}\BibitemShut {NoStop}%
\bibitem [{\citenamefont {Rudek}\ \emph {et~al.}(2012)\citenamefont {Rudek},
  \citenamefont {Son}, \citenamefont {Foucar}, \citenamefont {Epp},
  \citenamefont {Erk}, \citenamefont {Hartmann}, \citenamefont {Adolph},
  \citenamefont {Andritschke}, \citenamefont {Aquila}, \citenamefont {Berrah}
  \emph {et~al.}}]{Rudek_2012}%
  \BibitemOpen
  \bibfield  {author} {\bibinfo {author} {\bibfnamefont {B.}~\bibnamefont
  {Rudek}}, \bibinfo {author} {\bibfnamefont {S.-K.}\ \bibnamefont {Son}},
  \bibinfo {author} {\bibfnamefont {L.}~\bibnamefont {Foucar}}, \bibinfo
  {author} {\bibfnamefont {S.~W.}\ \bibnamefont {Epp}}, \bibinfo {author}
  {\bibfnamefont {B.}~\bibnamefont {Erk}}, \bibinfo {author} {\bibfnamefont
  {R.}~\bibnamefont {Hartmann}}, \bibinfo {author} {\bibfnamefont
  {M.}~\bibnamefont {Adolph}}, \bibinfo {author} {\bibfnamefont
  {R.}~\bibnamefont {Andritschke}}, \bibinfo {author} {\bibfnamefont
  {A.}~\bibnamefont {Aquila}}, \bibinfo {author} {\bibfnamefont
  {N.}~\bibnamefont {Berrah}},  \emph {et~al.},\ }\href@noop {} {\bibfield
  {journal} {\bibinfo  {journal} {Nature photonics}\ }\textbf {\bibinfo
  {volume} {6}},\ \bibinfo {pages} {858} (\bibinfo {year} {2012})}\BibitemShut
  {NoStop}%
\bibitem [{\citenamefont {Chen}\ \emph {et~al.}(2015)\citenamefont {Chen},
  \citenamefont {Pabst}, \citenamefont {Karamatskou},\ and\ \citenamefont
  {Santra}}]{Santra_Xe}%
  \BibitemOpen
  \bibfield  {author} {\bibinfo {author} {\bibfnamefont {Y.-J.}\ \bibnamefont
  {Chen}}, \bibinfo {author} {\bibfnamefont {S.}~\bibnamefont {Pabst}},
  \bibinfo {author} {\bibfnamefont {A.}~\bibnamefont {Karamatskou}}, \ and\
  \bibinfo {author} {\bibfnamefont {R.}~\bibnamefont {Santra}},\ }\href
  {\doibase 10.1103/PhysRevA.91.032503} {\bibfield  {journal} {\bibinfo
  {journal} {Phys. Rev. A}\ }\textbf {\bibinfo {volume} {91}},\ \bibinfo
  {pages} {032503} (\bibinfo {year} {2015})}\BibitemShut {NoStop}%
\bibitem [{\citenamefont {Makris}\ \emph {et~al.}(2009)\citenamefont {Makris},
  \citenamefont {Lambropoulos},\ and\ \citenamefont
  {Miheli\ifmmode~\check{c}\else \v{c}\fi{}}}]{Makris_21}%
  \BibitemOpen
  \bibfield  {author} {\bibinfo {author} {\bibfnamefont {M.~G.}\ \bibnamefont
  {Makris}}, \bibinfo {author} {\bibfnamefont {P.}~\bibnamefont
  {Lambropoulos}}, \ and\ \bibinfo {author} {\bibfnamefont {A.}~\bibnamefont
  {Miheli\ifmmode~\check{c}\else \v{c}\fi{}}},\ }\href {\doibase
  10.1103/PhysRevLett.102.033002} {\bibfield  {journal} {\bibinfo  {journal}
  {Phys. Rev. Lett.}\ }\textbf {\bibinfo {volume} {102}},\ \bibinfo {pages}
  {033002} (\bibinfo {year} {2009})}\BibitemShut {NoStop}%
\bibitem [{\citenamefont {Lambropoulos}\ \emph
  {et~al.}(2011{\natexlab{a}})\citenamefont {Lambropoulos}, \citenamefont
  {Papamihail},\ and\ \citenamefont {Decleva}}]{Lambropoulos_21}%
  \BibitemOpen
  \bibfield  {author} {\bibinfo {author} {\bibfnamefont {P.}~\bibnamefont
  {Lambropoulos}}, \bibinfo {author} {\bibfnamefont {K.~G.}\ \bibnamefont
  {Papamihail}}, \ and\ \bibinfo {author} {\bibfnamefont {P.}~\bibnamefont
  {Decleva}},\ }\href {\doibase 10.1088/0953-4075/44/17/175402} {\bibfield
  {journal} {\bibinfo  {journal} {Journal of Physics B: Atomic, Molecular and
  Optical Physics}\ }\textbf {\bibinfo {volume} {44}},\ \bibinfo {pages}
  {175402} (\bibinfo {year} {2011}{\natexlab{a}})}\BibitemShut {NoStop}%
\bibitem [{\citenamefont {Sorokin}\ \emph {et~al.}(2007)\citenamefont
  {Sorokin}, \citenamefont {Bobashev}, \citenamefont {Feigl}, \citenamefont
  {Tiedtke}, \citenamefont {Wabnitz},\ and\ \citenamefont
  {Richter}}]{Sorokin_21}%
  \BibitemOpen
  \bibfield  {author} {\bibinfo {author} {\bibfnamefont {A.~A.}\ \bibnamefont
  {Sorokin}}, \bibinfo {author} {\bibfnamefont {S.~V.}\ \bibnamefont
  {Bobashev}}, \bibinfo {author} {\bibfnamefont {T.}~\bibnamefont {Feigl}},
  \bibinfo {author} {\bibfnamefont {K.}~\bibnamefont {Tiedtke}}, \bibinfo
  {author} {\bibfnamefont {H.}~\bibnamefont {Wabnitz}}, \ and\ \bibinfo
  {author} {\bibfnamefont {M.}~\bibnamefont {Richter}},\ }\href {\doibase
  10.1103/PhysRevLett.99.213002} {\bibfield  {journal} {\bibinfo  {journal}
  {Phys. Rev. Lett.}\ }\textbf {\bibinfo {volume} {99}},\ \bibinfo {pages}
  {213002} (\bibinfo {year} {2007})}\BibitemShut {NoStop}%
\bibitem [{\citenamefont {Bergues}\ \emph {et~al.}(2018)\citenamefont
  {Bergues}, \citenamefont {Rivas}, \citenamefont {Weidman}, \citenamefont
  {Muschet}, \citenamefont {Helml}, \citenamefont {Guggenmos}, \citenamefont
  {Pervak}, \citenamefont {Kleineberg}, \citenamefont {Marcus}, \citenamefont
  {Kienberger}, \citenamefont {Charalambidis}, \citenamefont {Tzallas},
  \citenamefont {Schr\"{o}der}, \citenamefont {Krausz},\ and\ \citenamefont
  {Veisz}}]{Bergues}%
  \BibitemOpen
  \bibfield  {author} {\bibinfo {author} {\bibfnamefont {B.}~\bibnamefont
  {Bergues}}, \bibinfo {author} {\bibfnamefont {D.~E.}\ \bibnamefont {Rivas}},
  \bibinfo {author} {\bibfnamefont {M.}~\bibnamefont {Weidman}}, \bibinfo
  {author} {\bibfnamefont {A.~A.}\ \bibnamefont {Muschet}}, \bibinfo {author}
  {\bibfnamefont {W.}~\bibnamefont {Helml}}, \bibinfo {author} {\bibfnamefont
  {A.}~\bibnamefont {Guggenmos}}, \bibinfo {author} {\bibfnamefont
  {V.}~\bibnamefont {Pervak}}, \bibinfo {author} {\bibfnamefont
  {U.}~\bibnamefont {Kleineberg}}, \bibinfo {author} {\bibfnamefont
  {G.}~\bibnamefont {Marcus}}, \bibinfo {author} {\bibfnamefont
  {R.}~\bibnamefont {Kienberger}}, \bibinfo {author} {\bibfnamefont
  {D.}~\bibnamefont {Charalambidis}}, \bibinfo {author} {\bibfnamefont
  {P.}~\bibnamefont {Tzallas}}, \bibinfo {author} {\bibfnamefont
  {H.}~\bibnamefont {Schr\"{o}der}}, \bibinfo {author} {\bibfnamefont
  {F.}~\bibnamefont {Krausz}}, \ and\ \bibinfo {author} {\bibfnamefont
  {L.}~\bibnamefont {Veisz}},\ }\href@noop {} {\bibfield  {journal} {\bibinfo
  {journal} {Optica}\ }\textbf {\bibinfo {volume} {5}},\ \bibinfo {pages} {237}
  (\bibinfo {year} {2018})}\BibitemShut {NoStop}%
\bibitem [{\citenamefont {Cederbaum}\ \emph {et~al.}(1986)\citenamefont
  {Cederbaum}, \citenamefont {Tarantelli}, \citenamefont {Sgamellotti},\ and\
  \citenamefont {Schirmer}}]{MultipleCoreHoles1}%
  \BibitemOpen
  \bibfield  {author} {\bibinfo {author} {\bibfnamefont {L.~S.}\ \bibnamefont
  {Cederbaum}}, \bibinfo {author} {\bibfnamefont {F.}~\bibnamefont
  {Tarantelli}}, \bibinfo {author} {\bibfnamefont {A.}~\bibnamefont
  {Sgamellotti}}, \ and\ \bibinfo {author} {\bibfnamefont {J.}~\bibnamefont
  {Schirmer}},\ }\href {\doibase 10.1063/1.451432} {\bibfield  {journal}
  {\bibinfo  {journal} {The Journal of Chemical Physics}\ }\textbf {\bibinfo
  {volume} {85}},\ \bibinfo {pages} {6513} (\bibinfo {year}
  {1986})}\BibitemShut {NoStop}%
\bibitem [{\citenamefont {Tashiro}\ \emph {et~al.}(2010)\citenamefont
  {Tashiro}, \citenamefont {Ehara}, \citenamefont {Fukuzawa}, \citenamefont
  {Ueda}, \citenamefont {Buth}, \citenamefont {Kryzhevoi},\ and\ \citenamefont
  {Cederbaum}}]{MultipleCoreHoles2}%
  \BibitemOpen
  \bibfield  {author} {\bibinfo {author} {\bibfnamefont {M.}~\bibnamefont
  {Tashiro}}, \bibinfo {author} {\bibfnamefont {M.}~\bibnamefont {Ehara}},
  \bibinfo {author} {\bibfnamefont {H.}~\bibnamefont {Fukuzawa}}, \bibinfo
  {author} {\bibfnamefont {K.}~\bibnamefont {Ueda}}, \bibinfo {author}
  {\bibfnamefont {C.}~\bibnamefont {Buth}}, \bibinfo {author} {\bibfnamefont
  {N.~V.}\ \bibnamefont {Kryzhevoi}}, \ and\ \bibinfo {author} {\bibfnamefont
  {L.~S.}\ \bibnamefont {Cederbaum}},\ }\href {\doibase 10.1063/1.3408251}
  {\bibfield  {journal} {\bibinfo  {journal} {The Journal of Chemical Physics}\
  }\textbf {\bibinfo {volume} {132}},\ \bibinfo {pages} {184302} (\bibinfo
  {year} {2010})}\BibitemShut {NoStop}%
\bibitem [{\citenamefont {Lambropoulos}\ \emph
  {et~al.}(2011{\natexlab{b}})\citenamefont {Lambropoulos}, \citenamefont
  {Nikolopoulos},\ and\ \citenamefont {Papamihail}}]{Nikolopoulos_2011}%
  \BibitemOpen
  \bibfield  {author} {\bibinfo {author} {\bibfnamefont {P.}~\bibnamefont
  {Lambropoulos}}, \bibinfo {author} {\bibfnamefont {G.~M.}\ \bibnamefont
  {Nikolopoulos}}, \ and\ \bibinfo {author} {\bibfnamefont {K.~G.}\
  \bibnamefont {Papamihail}},\ }\href {\doibase 10.1103/PhysRevA.83.021407}
  {\bibfield  {journal} {\bibinfo  {journal} {Phys. Rev. A}\ }\textbf {\bibinfo
  {volume} {83}},\ \bibinfo {pages} {021407} (\bibinfo {year}
  {2011}{\natexlab{b}})}\BibitemShut {NoStop}%
\bibitem [{\citenamefont {Burhop}\ and\ \citenamefont {Asaad}(1972)}]{Auger}%
  \BibitemOpen
  \bibfield  {author} {\bibinfo {author} {\bibfnamefont {E.}~\bibnamefont
  {Burhop}}\ and\ \bibinfo {author} {\bibfnamefont {W.}~\bibnamefont {Asaad}}\
  }(\bibinfo  {publisher} {Academic Press},\ \bibinfo {year} {1972})\ pp.\
  \bibinfo {pages} {163 -- 284}\BibitemShut {NoStop}%
\bibitem [{\citenamefont {Fittinghoff}\ \emph {et~al.}(1992)\citenamefont
  {Fittinghoff}, \citenamefont {Bolton}, \citenamefont {Chang},\ and\
  \citenamefont {Kulander}}]{shake-off}%
  \BibitemOpen
  \bibfield  {author} {\bibinfo {author} {\bibfnamefont {D.~N.}\ \bibnamefont
  {Fittinghoff}}, \bibinfo {author} {\bibfnamefont {P.~R.}\ \bibnamefont
  {Bolton}}, \bibinfo {author} {\bibfnamefont {B.}~\bibnamefont {Chang}}, \
  and\ \bibinfo {author} {\bibfnamefont {K.~C.}\ \bibnamefont {Kulander}},\
  }\href {\doibase 10.1103/physrevlett.69.2642} {\enquote {\bibinfo {title}
  {{Observation of nonsequential double ionization of helium with optical
  tunneling}},}\ } (\bibinfo {year} {1992})\BibitemShut {NoStop}%
\bibitem [{\citenamefont {Rohringer}\ and\ \citenamefont
  {Santra}(2007)}]{Santra_RE_2007}%
  \BibitemOpen
  \bibfield  {author} {\bibinfo {author} {\bibfnamefont {N.}~\bibnamefont
  {Rohringer}}\ and\ \bibinfo {author} {\bibfnamefont {R.}~\bibnamefont
  {Santra}},\ }\href {\doibase 10.1103/PhysRevA.76.033416} {\bibfield
  {journal} {\bibinfo  {journal} {Phys. Rev. A}\ }\textbf {\bibinfo {volume}
  {76}},\ \bibinfo {pages} {033416} (\bibinfo {year} {2007})}\BibitemShut
  {NoStop}%
\bibitem [{\citenamefont {Wallis}\ \emph {et~al.}(2014)\citenamefont {Wallis},
  \citenamefont {Lodi},\ and\ \citenamefont {Emmanouilidou}}]{Wallis2014}%
  \BibitemOpen
  \bibfield  {author} {\bibinfo {author} {\bibfnamefont {A.~O.~G.}\
  \bibnamefont {Wallis}}, \bibinfo {author} {\bibfnamefont {L.}~\bibnamefont
  {Lodi}}, \ and\ \bibinfo {author} {\bibfnamefont {A.}~\bibnamefont
  {Emmanouilidou}},\ }\href {\doibase 10.1103/PhysRevA.89.063417} {\bibfield
  {journal} {\bibinfo  {journal} {Phys. Rev. A}\ }\textbf {\bibinfo {volume}
  {89}},\ \bibinfo {pages} {063417} (\bibinfo {year} {2014})}\BibitemShut
  {NoStop}%
\bibitem [{\citenamefont {Werner}\ \emph {et~al.}(2010)\citenamefont {Werner},
  \citenamefont {Knowles}, \citenamefont {Lindh}, \citenamefont {Manby},
  \citenamefont {{Sch\"{u}tz}} \emph {et~al.}}]{molpro}%
  \BibitemOpen
  \bibfield  {author} {\bibinfo {author} {\bibfnamefont {H.~J.}\ \bibnamefont
  {Werner}}, \bibinfo {author} {\bibfnamefont {P.~J.}\ \bibnamefont {Knowles}},
  \bibinfo {author} {\bibfnamefont {R.}~\bibnamefont {Lindh}}, \bibinfo
  {author} {\bibfnamefont {F.~R.}\ \bibnamefont {Manby}}, \bibinfo {author}
  {\bibfnamefont {M.}~\bibnamefont {{Sch\"{u}tz}}},  \emph {et~al.},\
  }\href@noop {} {\enquote {\bibinfo {title} {{MOLPRO}, a package of ab initio
  programs},}\ } (\bibinfo {year} {2010})\BibitemShut {NoStop}%
\bibitem [{\citenamefont {Martins}\ \emph {et~al.}(2013)\citenamefont
  {Martins}, \citenamefont {[de Souza]}, \citenamefont {Ceolin}, \citenamefont
  {Jorge}, \citenamefont {[de Berrêdo]},\ and\ \citenamefont {Campos}}]{AQZP}%
  \BibitemOpen
  \bibfield  {author} {\bibinfo {author} {\bibfnamefont {L.}~\bibnamefont
  {Martins}}, \bibinfo {author} {\bibfnamefont {F.}~\bibnamefont {[de Souza]}},
  \bibinfo {author} {\bibfnamefont {G.}~\bibnamefont {Ceolin}}, \bibinfo
  {author} {\bibfnamefont {F.}~\bibnamefont {Jorge}}, \bibinfo {author}
  {\bibfnamefont {R.}~\bibnamefont {[de Berrêdo]}}, \ and\ \bibinfo {author}
  {\bibfnamefont {C.}~\bibnamefont {Campos}},\ }\href@noop {} {\bibfield
  {journal} {\bibinfo  {journal} {Computational and Theoretical Chemistry}\
  }\textbf {\bibinfo {volume} {1013}},\ \bibinfo {pages} {62 } (\bibinfo {year}
  {2013})}\BibitemShut {NoStop}%
\bibitem [{\citenamefont {Herman}\ and\ \citenamefont
  {Skillman}(1963)}]{HermanSkillman1}%
  \BibitemOpen
  \bibfield  {author} {\bibinfo {author} {\bibfnamefont {F.}~\bibnamefont
  {Herman}}\ and\ \bibinfo {author} {\bibfnamefont {S.}~\bibnamefont
  {Skillman}},\ }\href@noop {} {\emph {\bibinfo {title} {Atomic structure
  calculations}}}\ (\bibinfo  {publisher} {Prentice-Hall, New Jersey},\
  \bibinfo {year} {1963})\BibitemShut {NoStop}%
\bibitem [{\citenamefont {Pauli}()}]{HermanSkillman2}%
  \BibitemOpen
  \bibfield  {author} {\bibinfo {author} {\bibfnamefont {M.~D.}\ \bibnamefont
  {Pauli}},\ }\href@noop {} {\enquote {\bibinfo {title} {{Herman-Skillman}
  program},}\ }\bibinfo {note}
  {Hermes.phys.uwm.edu/projects/elecstruct/elecstruct.html}\BibitemShut
  {NoStop}%
\bibitem [{\citenamefont {{Noumerov}}(1924)}]{Numerov}%
  \BibitemOpen
  \bibfield  {author} {\bibinfo {author} {\bibfnamefont {B.~V.}\ \bibnamefont
  {{Noumerov}}},\ }\href {\doibase 10.1093/mnras/84.8.592} {\bibfield
  {journal} {\bibinfo  {journal} {mnras}\ }\textbf {\bibinfo {volume} {84}},\
  \bibinfo {pages} {592} (\bibinfo {year} {1924})}\BibitemShut {NoStop}%
\bibitem [{\citenamefont {{Banks, Henry I. B.}}\ \emph
  {et~al.}(2020)\citenamefont {{Banks, Henry I. B.}}, \citenamefont
  {{Hadjipittas, Antonis}},\ and\ \citenamefont {{Emmanouilidou,
  Agapi}}}]{Hadjipittas_2019}%
  \BibitemOpen
  \bibfield  {author} {\bibinfo {author} {\bibnamefont {{Banks, Henry I. B.}}},
  \bibinfo {author} {\bibnamefont {{Hadjipittas, Antonis}}}, \ and\ \bibinfo
  {author} {\bibnamefont {{Emmanouilidou, Agapi}}},\ }\href {\doibase
  10.1140/epjd/e2020-100416-6} {\bibfield  {journal} {\bibinfo  {journal} {Eur.
  Phys. J. D}\ }\textbf {\bibinfo {volume} {74}},\ \bibinfo {pages} {98}
  (\bibinfo {year} {2020})}\BibitemShut {NoStop}%
\bibitem [{\citenamefont {Sakurai}(1994)}]{Sakurai}%
  \BibitemOpen
  \bibfield  {author} {\bibinfo {author} {\bibfnamefont {J.~J.}\ \bibnamefont
  {Sakurai}},\ }\href@noop {} {\emph {\bibinfo {title} {Modern Quantum
  Mechanics}}}\ (\bibinfo  {publisher} {Addison-Wesley},\ \bibinfo {year}
  {1994})\BibitemShut {NoStop}%
\bibitem [{\citenamefont {Edmonds}(1960)}]{Weigner_3j}%
  \BibitemOpen
  \bibfield  {author} {\bibinfo {author} {\bibfnamefont {A.~R.}\ \bibnamefont
  {Edmonds}},\ }\href@noop {} {\emph {\bibinfo {title} {Angular Momentum in
  Quantum Mechanics}}}\ (\bibinfo  {publisher} {Princeton University Press},\
  \bibinfo {year} {1960})\BibitemShut {NoStop}%
\bibitem [{\citenamefont {Becker}\ \emph {et~al.}(1989)\citenamefont {Becker},
  \citenamefont {Szostak}, \citenamefont {Kerkhoff}, \citenamefont {Kupsch},
  \citenamefont {Langer}, \citenamefont {Wehlitz}, \citenamefont {Yagishita},\
  and\ \citenamefont {Hayaishi}}]{Becker}%
  \BibitemOpen
  \bibfield  {author} {\bibinfo {author} {\bibfnamefont {U.}~\bibnamefont
  {Becker}}, \bibinfo {author} {\bibfnamefont {D.}~\bibnamefont {Szostak}},
  \bibinfo {author} {\bibfnamefont {H.~G.}\ \bibnamefont {Kerkhoff}}, \bibinfo
  {author} {\bibfnamefont {M.}~\bibnamefont {Kupsch}}, \bibinfo {author}
  {\bibfnamefont {B.}~\bibnamefont {Langer}}, \bibinfo {author} {\bibfnamefont
  {R.}~\bibnamefont {Wehlitz}}, \bibinfo {author} {\bibfnamefont
  {A.}~\bibnamefont {Yagishita}}, \ and\ \bibinfo {author} {\bibfnamefont
  {T.}~\bibnamefont {Hayaishi}},\ }\href {\doibase 10.1103/PhysRevA.39.3902}
  {\bibfield  {journal} {\bibinfo  {journal} {Phys. Rev. A}\ }\textbf {\bibinfo
  {volume} {39}},\ \bibinfo {pages} {3902} (\bibinfo {year}
  {1989})}\BibitemShut {NoStop}%
\bibitem [{\citenamefont {Holland}\ \emph {et~al.}(1979)\citenamefont
  {Holland}, \citenamefont {Codling}, \citenamefont {Marr},\ and\ \citenamefont
  {West}}]{Holland_1979}%
  \BibitemOpen
  \bibfield  {author} {\bibinfo {author} {\bibfnamefont {D.~M.~P.}\
  \bibnamefont {Holland}}, \bibinfo {author} {\bibfnamefont {K.}~\bibnamefont
  {Codling}}, \bibinfo {author} {\bibfnamefont {G.~V.}\ \bibnamefont {Marr}}, \
  and\ \bibinfo {author} {\bibfnamefont {J.~B.}\ \bibnamefont {West}},\ }\href
  {\doibase 10.1088/0022-3700/12/15/008} {\bibfield  {journal} {\bibinfo
  {journal} {Journal of Physics B: Atomic and Molecular Physics}\ }\textbf
  {\bibinfo {volume} {12}},\ \bibinfo {pages} {2465} (\bibinfo {year}
  {1979})}\BibitemShut {NoStop}%
\bibitem [{\citenamefont {Yeh}\ and\ \citenamefont {Lindau}(1985)}]{Yeh}%
  \BibitemOpen
  \bibfield  {author} {\bibinfo {author} {\bibfnamefont {J.}~\bibnamefont
  {Yeh}}\ and\ \bibinfo {author} {\bibfnamefont {I.}~\bibnamefont {Lindau}},\
  }\href@noop {} {\bibfield  {journal} {\bibinfo  {journal} {Atomic Data and
  Nuclear Data Tables}\ }\textbf {\bibinfo {volume} {32}},\ \bibinfo {pages} {1
  } (\bibinfo {year} {1985})}\BibitemShut {NoStop}%
\bibitem [{\citenamefont {{Amusia}}\ \emph {et~al.}(1973)\citenamefont
  {{Amusia}}, \citenamefont {{Chernysheva}},\ and\ \citenamefont
  {{Ivanov}}}]{Amusia_1973}%
  \BibitemOpen
  \bibfield  {author} {\bibinfo {author} {\bibfnamefont {M.~Y.}\ \bibnamefont
  {{Amusia}}}, \bibinfo {author} {\bibfnamefont {L.~V.}\ \bibnamefont
  {{Chernysheva}}}, \ and\ \bibinfo {author} {\bibfnamefont {V.~K.}\
  \bibnamefont {{Ivanov}}},\ }\href {\doibase 10.1016/0375-9601(73)90289-2}
  {\bibfield  {journal} {\bibinfo  {journal} {Physics Letters A}\ }\textbf
  {\bibinfo {volume} {43}},\ \bibinfo {pages} {243} (\bibinfo {year}
  {1973})}\BibitemShut {NoStop}%
\bibitem [{\citenamefont {Lambropoulos}\ and\ \citenamefont
  {Tang}(1987)}]{Lambropoulos_Scaling}%
  \BibitemOpen
  \bibfield  {author} {\bibinfo {author} {\bibfnamefont {P.}~\bibnamefont
  {Lambropoulos}}\ and\ \bibinfo {author} {\bibfnamefont {X.}~\bibnamefont
  {Tang}},\ }\href {\doibase 10.1364/JOSAB.4.000821} {\bibfield  {journal}
  {\bibinfo  {journal} {J. Opt. Soc. Am. B}\ }\textbf {\bibinfo {volume} {4}},\
  \bibinfo {pages} {821} (\bibinfo {year} {1987})}\BibitemShut {NoStop}%
\bibitem [{\citenamefont {Pauli}(2000)}]{Fermi}%
  \BibitemOpen
  \bibfield  {author} {\bibinfo {author} {\bibfnamefont {W.}~\bibnamefont
  {Pauli}},\ }\href@noop {} {\emph {\bibinfo {title} {Wave Mechanics: Volume 5
  of Pauli Lectures on Physics}}}\ (\bibinfo  {publisher} {Wiley},\ \bibinfo
  {year} {2000})\ pp.\ \bibinfo {pages} {150--151}\BibitemShut {NoStop}%
\bibitem [{\citenamefont {Banks}\ \emph {et~al.}(2017)\citenamefont {Banks},
  \citenamefont {Little}, \citenamefont {Tennyson},\ and\ \citenamefont
  {Emmanouilidou}}]{Emmanoulidou_Banks_Tennyson_Auger}%
  \BibitemOpen
  \bibfield  {author} {\bibinfo {author} {\bibfnamefont {H.~I.~B.}\
  \bibnamefont {Banks}}, \bibinfo {author} {\bibfnamefont {D.~A.}\ \bibnamefont
  {Little}}, \bibinfo {author} {\bibfnamefont {J.}~\bibnamefont {Tennyson}}, \
  and\ \bibinfo {author} {\bibfnamefont {A.}~\bibnamefont {Emmanouilidou}},\
  }\href {\doibase 10.1039/C7CP02345F} {\bibfield  {journal} {\bibinfo
  {journal} {Phys. Chem. Chem. Phys.}\ }\textbf {\bibinfo {volume} {19}},\
  \bibinfo {pages} {19794} (\bibinfo {year} {2017})}\BibitemShut {NoStop}%
\bibitem [{\citenamefont {{{\AA}}berg}(1969)}]{Aberg1969}%
  \BibitemOpen
  \bibfield  {author} {\bibinfo {author} {\bibfnamefont {T.}~\bibnamefont
  {{{\AA}}berg}},\ }\href@noop {} {\bibfield  {journal} {\bibinfo  {journal}
  {Ann. Acad. Sci. Fenn., Ser A VI}\ }\textbf {\bibinfo {volume} {308}},\
  \bibinfo {pages} {1} (\bibinfo {year} {1969})}\BibitemShut {NoStop}%
\bibitem [{\citenamefont {Carlson}\ and\ \citenamefont
  {Nestor}(1973)}]{Carlson1973}%
  \BibitemOpen
  \bibfield  {author} {\bibinfo {author} {\bibfnamefont {T.~A.}\ \bibnamefont
  {Carlson}}\ and\ \bibinfo {author} {\bibfnamefont {C.~W.}\ \bibnamefont
  {Nestor}},\ }\href {\doibase 10.1103/PhysRevA.8.2887} {\bibfield  {journal}
  {\bibinfo  {journal} {Phys. Rev. A}\ }\textbf {\bibinfo {volume} {8}},\
  \bibinfo {pages} {2887} (\bibinfo {year} {1973})}\BibitemShut {NoStop}%
\end{thebibliography}%
\bibliographystyle{apsrev4-1}

\end{document}